\documentclass[11pt,a4paper]{article} 
\usepackage[title]{appendix}
\usepackage{epsfig,amssymb,amsmath,latexsym} 
\usepackage{jcappubmy} 
\usepackage{hyperref}
\usepackage{pifont} 
\usepackage{hyperref} 
\usepackage{float}
\usepackage{caption}
\usepackage{subfigure}
\usepackage{color} 
\definecolor{nicered}{rgb}{0.7,0.1,0.1} 
\definecolor{nicegreen}{rgb}{0.1,0.5,0.1}

\newcommand{\ar}{\hat{r}}
\newcommand{\be}{\begin{equation}} 
\newcommand{\ee}{\end{equation}} 
\newcommand{\bea}{\begin{eqnarray}} 
\newcommand{\eea}{\end{eqnarray}}

\renewcommand\l{\lambda} 
\renewcommand\o{\omega} 
\renewcommand\a{\alpha} 
 
\renewcommand\b{\beta} 
\renewcommand\k{\kappa} 
\newcommand\e{\epsilon} 
\newcommand\m{\mu} 
\newcommand\n{\nu} 
\newcommand\g{\gamma}

\renewcommand\l{\ensuremath{\lambda}}

\newcommand\ba{\begin{array}} 
\newcommand\ea{\end{array}}

 \title{Inhomogeneous   model   with a space dependent Cosmological Constant } 
 
\author[]{ {\Huge } D. Comelli }
\affiliation[]{INFN  Ferrara Division, via Saragat  1, Ferrara, 44122, Italy}
 
\emailAdd{comelli@fe.infn.it}

\date{\small \today} 
 
\abstract{  
  We analyse an inhomogeneous cosmological model featuring a
   spherically symmetric bubble solution induced by 
   a unified single perfect fluid,   comprising    spatially dependent  Dark Energy (with $w=-1$) and   Dark Matter (with $w=0$) components.
    We impose an Hubble profile that matches the Planck value at early times ($H_0=67.4\pm0.5$ Km/s\,Mpc) and  the local value
   (${\cal H}_0=73.52\pm 1.62$ Km/s\,Mpc).
  We explicitly derive  perturbative solutions in two distinct regimes: one expanded around the center (for small $r\,{\cal H}_0\ll1$) and the other expanded around a  homogeneous FRW universe. 
    In both cases, we compute the  cosmographic parameters, 
    redshift profiles for the Hubbles expansion rates, and effective equations of state. Furthermore, we investigate the  redshift drift behaviour  extended to a Lema$\hat{\rm i}$tre metric.
  }

\begin{document}

\maketitle 
 
 \section{Introduction}
 With the discovery of an accelerated expansion of the universe, cosmology entered into the Dark Energy  (DE) era \cite{SupernovaCosmologyProject:1998vns}.
 The introduction of a Cosmological Constant (CC) inside the Einstein eqs is the simplest resolution to explain the main features of the present data. However, with the advent of more observations some tensions have energed. 
    In particular,  the $H_0$-tension problem arises because, while the CMB Planck data yields a value for the Hubble constant $H_0=67.4\pm0.5\,Km/s/Mpc$ within the $\Lambda CDM$ model \cite{Planck}, 
    local measurements favour higher values, such as  ${\cal H}_0=73.52\pm1.62\,Km/s/Mpc$ \cite{Riess}.\footnote{Here, $H_0$  refers to the  Hubble constant from Planck, while   ${\cal H}_0$ denotes  the nearby Hubble constant.}
  The explanation of this discrepancy has led to a significant number of studies  (see \cite{DiValentino:2021izs} and  \cite{Hu:2023jqc} as general reviews).
 In a recent paper \cite{Comelli:2023otq}, we discussed an alternative proposal that can be categorized, following the classification   of \cite{DiValentino:2021izs}, as a solution under the chapter  {\it Local Inhomogeneity}, representing a viable    alternative for late-time evolution.
 In this study, we move away from the homogeneous space-time typically characterized by FRW backgrounds \cite{Aluri:2022hzs} and instead focus on a spherical void model \cite{vanElst:1995eg} generated by a unified single perfect fluid that behaves as both Dark Matter (DM) and Dark Energy (DE). While spherical models containing only DM belong to the class of Lemaitre-Tolman-Bondi (LTB) models \cite{Enqvist:2007vb}, \cite{Bellido}, the inclusion of a DE (radially dependent) component has not been previously studied.
  The main geometrical difference from the LTB models is the presence of a spatial gradient in the pressure (\cite{Bolejko}, \cite{Moradi:2013gf}, \cite{Hellaby}), which induces acceleration for comoving observers.\footnote{NB: the inhomogeneous pressure Stephani model is a spherically symmetric solution of the Einstein equations for shear-free  perfect fluids ($\Sigma=0$) (\ref{kin}) with a density and pressure of the form $\rho=\rho(t)$ and $p=p(t,r)$) , but this  is not our case (\ref{consd}).}
  In a series of papers (\cite{Comelli:2023otq}, \cite{Celoria}), we studied  a Lagrangian model
   generating a perfect fluid  where the   DE component  is linked to the thermodynamic intrinsic entropy of the fluid, conserved over time but space-dependent.
  In the present work, we omit the  detailed theoretical structure of such a models and focus on a
  single Perfect Fluid (PF) system, where the
  energy density $\rho(t,r)$ is the sum of a space-dependent CC $\Lambda(r)$ and a DM component proportional to the conserved density $n(t,r)$.  The pressure $p(r)$ has only the negative DE   contribution:
  \be\label{consd}
    \rho(t,r)=\Lambda(r)+m\,n(t,r),\quad p(r)=-\Lambda(r)  
    \ee
     where $m$ is a mass parameter that we can absorb into $n$, so we set $m=1$.
 \\
 We stress that  such a system is a single fluid,  meaning that   its energy-momentum tensor (EMT)
  follows a single conservation law that unifies the two components (DM and DE) (see (\ref{emt})).
  In most cases, the EMTs of  DM and DE follow  separate conservation laws. 
 We   study  this simple system by solving its evolution equations  using two different approximations.
 \\
  In the first approximation (Chapter \ref{near}), we 
 expand the  Einstein eqs for small  values of $r\,{\cal H}_0$ values and we obtain solutions for the nearby central region.
 \\
 In the second approximation (Chapter \ref{pertf}),  
 we expand around a FRW solution, i.e.,  we treat small    deviations (dependent on $r$)
  from the constant values of the CC energy density   
 ($\rho_\Lambda=const$) and from the DM density  ($a(t)^3\,n_{DM}=const$) (where scale factor $a(t)$ satisfies eq. (\ref{HCDM})).
 \\
 Finally, in Chapter \ref{geolc}, we analyse the past light   geodesic eqs, defining various observables in the redshift space. 
 \\
  Moreover, in Chapter \ref{dri}, we study  the bubble's imprint  on the   redshift drift, improving its definition  within the context of a  Lema$\hat{\rm i}$tre metric.
 \\
 A drawback of this approach is the breaking of the homogeneity.
 The presence of a center implies that off-center observers   would perceive  an anisotropic universe \cite{Alnes}.  For example, the CMB dipole  provides  stringent bounds, allowing at most a 
  displacement of 15 Mpc   from the center of the underdense bubble \cite{Bellido}, \cite{Alnes},  \cite {Cusin:2016kqx}, \cite{Biswas:2010xm} (a larger  estimate is possible once our peculiar motion cancels part of the dipole).
  In our case, following \cite{Enqvist:2007vb}, we reevaluate  the fine tuning required, taking into account the  peculiar velocity of an observer at the comoving coordinate $r = d$  relative to the symmetry center: 
  \bea
  v(d)=d\;\left( {\cal H}_0- {\cal H}_0(d) \right),\qquad
   {\cal H}_0= { H}_0+\Delta H_0,\;\; {\cal H}_0(r)={H}_0+\Delta H_0\,e^{-r^2/\Delta^2}
  \eea 
  see (\ref{lp}, \ref{leh1}), this uses the deviation from the Hubble law. Taking as an order of magnitude the CMB dipole velocity $v\sim 10^{-3}$, we obtain
    \bea
  d\sim \sqrt[3]{ \frac{\Delta^2}{\Delta H_0}\; v } \sim 3.67\;\left(\frac{\Delta}{Mpc}\right)^{2/3} Mpc \sim 990\;\Re^{2/3}\,Mpc,\qquad \Delta \equiv \frac{\Re}{H_0}
  \eea
 This is a non-negligible fraction of the inhomogeneity, with $d/\Delta\sim 0.48$, improving the constraints from  fine-tuning considerations. A similar result is obtained in \cite{Enqvist:2007vb}, where the Hubble profile  is $H_0+\Delta H_0\,e^{-r/\Delta}$. 
    \section{Lema\texorpdfstring{$\hat{\rm i}$}{Lgg}tre spacetime metric}
 A generic spherically symmetric spacetime can be described, using comoving coordinate, by the following Lema$\hat{\rm i}$tre metric  \cite{Hellaby}, \cite{Lemaitre}, \cite{hellab}
 \be\label{g}
ds^2=-F^2(t,r)\;dt^2+X^2(t,r)\;dr^2+R^2(t,r)\;(\;d\theta^2+\sin(\theta)\;d\phi^2)
\ee
 The matter flow has 4-velocity  $u^\a=(1/F,\;0,\;0,\;0)$, while the orthogonal   space like 4-vector is $v^\a=(0,\;1/X,\;0,\;0)$.
There are generically  two covariant derivatives for a scalar function $f(t,r)$, one along the  fluid flow lines $u^\a$ and one along the preferred space-like direction $v^\a$:
\be\label{Dir}
\frac{d f}{d\tau}\equiv\dot f\equiv u^\a\nabla_\a \,f= \frac{f^{(1,0)}}{F},\qquad  
\frac{d f}{d\mathfrak{r}}\equiv\hat{f} \equiv v^\a\nabla_\a \,f=\frac{f^{(0,1)}}{X}
\ee 
where $d\tau=F\,dt$ and $d\mathfrak{r}=X\,dr$.
Note also that in many articles   the $X(t,r)$ function is parametrised as
\be
X(t,r)\equiv \frac{R^{(0,1)}(t,r)}{\sqrt{1+E(t,r)}}\;\to\; E =\hat R^2-1
\ee
where the  $E$ function ($E>-1$) is the curvature parameter \cite{Lasky:2010vn}.
The geometrical quantities   derived from the gradient of the 4-velocity matter field  for a  Locally Rotationally  Symmetric (LRS) spacetime  are presented in Appendix (\ref{AppA}). The expressions for the acceleration ${\cal A}$, shear $\Sigma$, and   expansion $\theta$, using the metric (\ref{g}), result
\bea\label{kin}
&& {\cal A}=\frac{\hat F}{F},
   \qquad
    \Sigma=\frac{2}{3}\left(\frac{\dot X}{X}-\frac{\dot R}{R}\right)
 ,\qquad
   \theta = \left(\frac{\dot X}{X}+2\;\frac{\dot R}{R}\right)
\eea
The energy momentum tensor (EMT) of our system is that of a PF
\be
T_{\m\n}=(p+\rho)\;u_\m\;u_\n+p\;g_{\m\n}
\ee
where density and pressure are given by
\be
\rho(t,r)=\Lambda(r)+ n(t,r),\quad p(r)=-\Lambda(r)\label{rho}
\ee
The key point about such a fluid is the fact that the EMT is \underline{not} the sum of two separately conserved   components
 \bea\label{emt}
 &&T^{(DE)}_{\m\n}=\Lambda\,g_{\m\n},\quad T^{(DM)}_{\m\n}=n\,u_m\,u_\n, 
\quad \nabla^\m T^{(DE)}_{\m\n}\neq 0,\;\;\nabla^\m T^{(DM)}_{\m\n}\neq 0,
 \\&& {\rm but}\quad \nabla^\m \left(T^{(DE)}_{\m\n}+T^{(DM)}_{\m\n}\right)= 0,\nonumber
 \eea
 This implies that DE and DM are strongly coupled, forming a single fluid component.
 \\
Our goal  is to derive  a closed equation for the $R(t,r)$ function. 
Energy  density conservation   is linked to the   conservation of density $n$ and to the time constancy 
 of the CC
\bea
&&\dot n+\theta\;n=0\quad \oplus \quad \dot\Lambda=0 \quad\leftrightarrow \quad\dot \rho+\theta\;(\rho+p)=0 
\eea
The   eq. for the density can be easily integrated
\be\label{dens}
n(t,r)=\frac{{\bf n}(r)}{V[t,r]}=\frac{{\bf n}(r)}{X(t,r)\;R(t,r)^2},\qquad \theta=\frac{\dot V}{V}
\ee
where    ${\bf n}(r)$ is an arbitrary spatial function (boundary condition). 
The second eq.  for the CC gives $\Lambda\equiv \Lambda(r)$ (again a spatial  boundary condition). 
Finally, the space component of the EMT conservation, $h^\n_\m\nabla^\l T_{\l \n}=0$, results  
\bea \label{eqA}
h^\n_\m\nabla_\n p\equiv D_\m p=-(\rho+p)\;{\cal A}_\m\to D_\m \Lambda= n\;{\cal A}_\m,\qquad h^\n_\m=\delta^\n_\m+u^\n\,u_\m
\eea
where we see that a space-dependent CC induces an acceleration ${\cal A}_\m$ for the comoving observers.\footnote{Note from eq. (\ref{eqA}) that a non-trivial CC is possible only in the presence of a DM contribution, i.e., when  $n\neq 0$.
 Without the DM component, we are restricted to Lema$\hat{\rm i}$tre  $\Lambda=const$ solutions. This is why  the naive limit 
$\Omega_m\to 0$ in some eqs is not smooth (see for example (\ref{fxr},\ref{xfr},\ref{rfx})).
} 
Now we introduce the Misner-Sharp mass   \cite{Lemaitre} 
\be\label{eqdM}
M(t,r)\equiv
   \frac{ R}{2}\;
   \left(1+  \dot R^2  -  \hat R^2  \right)=\frac{ R}{2}\;
   \left(   \dot R^2  -  E  \right)
\ee which, in the Newtonian limit, represents  the mass inside a shell of radial coordinate $r$.
Using the $M(t,r)$ function,  the Einstein eqs read
\be\label{eqpr}
\rho=\Lambda+n=
2\;\frac{\hat M }{R^2\;\hat R },\qquad 
-p= \Lambda=
 2\;\frac{\dot M }{R^2\;\dot R }
\ee
Note that a curvature singularity ($\rho\to \infty$), called {\it shell crossing}, occurs
 for $\hat R\to 0$, unless   $\hat M\to 0$. This is not the case for our functional choices.
The   pressure equation can be  integrated, yielding
\be\label{eqM}
 M= \frac{1}{6}\;\left(\Lambda  \;R^3 \;+\mathbf{m}(r) \right)
\ee
where   $\mathbf{m}\equiv\mathbf{m}(r)$ is another   spatial  boundary condition.
Inserting  (\ref{eqM}) in (\ref{eqdM}),  we define the {\it transverse Hubble expansion rate }  (i.e.  the rate corresponding to the angular directions)    ${\cal H}_{\perp}$, which is related to the time derivative of  $R$  as\footnote{The mass dimensions of the various functions are
$\,[R]=[{\bf m}]=[M]=-1,\,[F]=[X]=[{\bf n}]=[E]=[{\bf E}]=0,\;[\Lambda]=[n]=[p]=[\rho]=+2$.}:
\bea\label{eqY}
{\cal H}_{\perp}^2\equiv \frac{\theta}{3}-\frac{\Sigma}{2}=\left(\frac{\dot R}{R}\right)^2&=&\left(\frac{R^{(1,0)}}{F\;R}\right)^2=\frac{1}{3}\,
\left( \Lambda  +  \frac{\bf m}{ R^3}  +   \frac{3\;E}{R^2}  \right)
\eea
while the radial derivative of $R$, using the expression for the energy density (\ref{eqpr}) and the density (\ref{dens}), can be written  as
 \bea\label{eqYh}
 \hat{R}=\frac{X}{3\;{\bf n}}\;\left( \hat{\bf m}+\hat{\Lambda}\;R^3\right)=\frac{1}{3\;{\bf n}}\;\left(    {\bf m}'+{\Lambda}'\;R^3\right)\equiv \sqrt{E(t,r)+1}
 \eea
The same expression can be used also to  relate $n$ and $\bf{m}$ as
\be\label{mn}
\hat{\bf{m}}=n\;R^2\;\left(3\;\hat{R}-\frac{\hat F}{F}\;R \right)
\ee
At this point, we have three arbitrary space-dependent functions ${\bf n},\;{\bf m},\;{\Lambda}$
and the following system of eqs for $F$, $R$, and  $X$ :
\bea
&&\frac{\hat F}{F}=\frac{3\,\hat{\Lambda}\;R^2\,\hat R}
{\hat{\bf m}+\hat{\Lambda}\,R^3}\to \label{eqF}
\boxed{
 \frac{F^{(0,1)}}{F}= \frac{3\;\Lambda '\; R^2 \;R^{(0,1)} }{ \mathbf{m}'+  \Lambda '\;R^3
 }}
 \\ 
 && {\cal H}_{\perp}^2\equiv \left(\frac{\dot R}{R}\right)^2 = \boxed{\left(\frac{R^{(1,0)}}{F\;R}\right)^2=\frac{1}{3}\;
\left( \!\!\underbrace{\Lambda }_{DE} \!\!+ \underbrace{ \frac{\bf m}{ R^3}}_{DM} \!\! +  \underbrace{\frac{3\;E}{R^2}}_{Curvature} \right)}
\\\label{eqX}
&& X=\frac{3\,{\bf n}\;R^{(0,1)}}{\Lambda'\,R^3+{\bf m}'}
\eea
We can introduce also the {\it longitudinal Hubble expansion rate} ${\cal H}_{\parallel}$ (i.e. the observed radial Hubble rate connected to the line of sight expansion) \cite{Jimenez} given by \footnote{The local Hubble rate related to the congruence $u^\a$ is given by ${\cal H}_\theta\equiv\nabla^a u_a=\frac{\theta}{3}  =\frac{1}{3}({\cal H}_{\parallel}+2\,{\cal H}_{\perp})+e^\a\,{\cal A}_\a$. 
\\ In LTB models is introduced also a radial Hubble parameter defined as $\partial_{tr}R/\partial_rR$.}
\be\label{hl}
{\cal H}_{\parallel}\equiv \frac{\theta}{3}+\Sigma-{\cal A}=\frac{\dot X}{X}-\frac{\hat{F}}{F}=
\frac{X^{(1,0)}-F^{(0,1)}}{F\,X}
\ee
Finally, we have also the {\it volume expansion rate} \cite{vanElst:1995eg} given by
\be\label{Htheta}
{\cal H}_\theta=\frac{\theta}{3}=\frac{{\cal H}_\parallel+2\,{\cal H}_\perp+{\cal A}}{3}=\frac{1}{3}\left(\frac{\dot X}{X}+2\,\frac{\dot R}{R}\right)=\frac{1}{3\,F}\left(\frac{  X^{(1,0)} }{X}+2\,\frac{R^{(1,0)}}{R}\right)
\ee
Using the kinematical variable (\ref{kin}), we observe that  the shear $\Sigma$ and the acceleration ${\cal A}$ are 
related to  different Hubble expansion rates  by
\bea\label{sa}
\Sigma=2\,({\cal H}_\theta-{\cal H}_\perp),\quad {\cal A}=3\,{\cal H}_\theta-2\,{\cal H}_\perp-{\cal H}_\parallel
\eea
To simplify the computations we define the space combination
 \be\label{fE}
      {\bf E}(r)\equiv \left(\frac{{\bf m}'}{3\,{\bf n}}\right)^2-1
      \ee
      so that our free spatial functions are now ${ \Lambda},\,{\bf m},\,{\bf E}$ (instead of $\bf n$).
      \\
      The limit 
       to the LTB and FRW cases is given by:
      \begin{itemize}
      \item {\it LTB}: From eq. (\ref{eqF}), we have $\hat F =0$, allowing us to set $F=1$.
   In eq. (\ref{eqYh}),
     when $\Lambda'\to0$ (i.e.  when $\Lambda$ is constant in both space and time), we obtain
      the time independent limit   $E(t,r)\to{\bf E}(r),$ so that eq. (\ref{eqX}) becomes
     \be\label{ltbx}
     \left. X \right |_{LTB}=\frac{3\,{\bf n}\,R^{(0,1)}}{{\bf m}'}=\frac{ R^{(0,1)}}{\sqrt{1+{\bf E}(r)}}
     \ee
     This corresponds exactly to the LTB parametrisation with ${\bf E}(r)>-1$ \cite{Enqvist:2007vb}.
     
\item {\it FRW }:
To achieve a spatially homogeneous universe, we require the following conditions:
\be\label{frwL}
  \!\!\! \!\!\! \Lambda(r)=H_0^2\;\Omega_{\Lambda},\;\; 
{\bf m}(r)=H_0^2\;\Omega_{m}\;r^3,
\;\;
{\bf n}(r)=\frac{{\bf m}'(r)}{3}=H_0^2\;\Omega_{m}\;r^2 \;{\rm and}\; {\bf E}(r)=0
\ee
For the functions of the metric, we have $\hat R=1,\;\hat F=\hat X= E=0$, so that  only eq. (\ref{eqY}) remains.
\end{itemize}
Note that the functions ${\cal H}_{\parallel}$ (\ref{hl}), ${\cal H}_{\perp}$ (\ref{eqY}), and ${\cal H}_{\theta}$  (\ref{Htheta}) in a general Lema$\hat{\rm i}$tre spacetime do not exhibit
 any functional correlation. However    in the LTB and FRW cases, we observe specific relationships 
 between these functions 
\bea\label{Hs}
&&{\rm  FRW:}\;\;  {\cal H}_{\parallel}={\cal H}_{\perp}={\cal H}_{\theta},\\
&&{\rm  LTB:}\;\;  {\cal H}_{\parallel}=\frac{R^{(1,1)}}{R^{(0,1)}}={\cal H}_{\perp}\,\left(1+\frac{\partial_r\log {\cal H}_{\perp}}{\partial_r\log R}\right),\qquad {\cal H}_{\theta}=\frac{{\cal H}_\parallel+2\,{\cal H}_\perp}{3}
\nonumber
\eea
      
      \section{Geodesic light equations} \label{geolc}
The determination of the light paths in a given spacetime are given, in the eikonal approximation, by the irrotational null geodesic eqs   \cite{Synge}
\be\label{eql}
\frac{d k^\a}{d\l}=k^\b\,\nabla_\b\,k^\a=0, \quad k^\a\,k_\a=0,\quad \nabla_{[\a}k_{\b]}=0,\quad k^\m\equiv\frac{d x^\m(\l)}{d\l}
\ee
where $k^\m$ is the tangent vector to the light path  $x^\m(\l)$ (a null geodesic) and $\l$ is an affine 
parameter. The $k^\a$ vector can be further parametrised as \cite{deSwardt:2010nf},\cite{Umeh:2014ana} 
\footnote{For a generic function $f$ we have already defined the directional derivatives $\dot f$
 and $\hat f$ from eq. (\ref{Dir}).
Then we can write the directional derivative along a light ray as
\be\label{dir}
f'\equiv \frac{d f}{d\lambda}=k^\mu\nabla_\mu f=E\,(u^\m+\k\;v^\m+\k^\m)\,\nabla_\m\,f
\ee
where we used the decomposition (\ref{grk}) of $k^\mu$.
In a LRS space, the propagation of a generic scalar function $f(t,r)$ along a {\it radial} light geodesic vector, where $\k=-1$ and $\k^\mu=0$, can be written as
\be\label{dirf}
f'=E\;(\dot f-\hat f)   
\ee}
\bea\label{deck}
k^\a=E\,(u^\a+d^\a),\quad d^\a\,u_\a=0,\quad d^\a\,d_\a=1
\eea
where the vector $d^\m$ is the spatial direction of propagation while $E=-u^\a k_\a$ is the energy of the photon seen by the 
committing observer $u^\a$. 
We consider an observer ${\cal O}$ located at the center of symmetry, at $r=0$, 
connected by an ingoing radial null geodesic to a   comoving source ${\cal E}$ at $r$.
The redshift $z$ is\footnote{We can normalise $\l$  such  that $(E_{{\cal O} }=-u\cdot k)_{{\cal O} }=1$.} 
related to the ratio of the energy at the emission point to the  the  energy at the observation point.
In general we can express it as:
\bea\label{defz}
1+z=\frac{(u\cdot k)_{ {\cal E}}}{(u\cdot k)_{{\cal O} }}=\frac{E_{ {\cal E}}}{E_{{\cal O} }},\quad 
(u\cdot k)_{{\cal O},{\cal E}}=\left.\frac{k^0(t,r)}{F(t,r)}\right |_{{\cal O},{\cal E}}\!\!\!\!, \;\; (1+z(\l))=F(t(\l),r(\l))\;t'(\l)
\eea
Note that while $z$ is an observable quantity, the $\l$ parameter is not. 
To compare with observations, we need the relationship  between the two, which is 
 given by the following differential eq:
\bea\nonumber
\!\!\!\!\!\!\!\!\!\!\!\!\frac{dE}{d\l}&=&-\frac{d(u^\a\,k_\a)}{d\l}=
-k^\b\,k^\a\nabla_\b\,u_\a
=-E^2\;\left(\frac{\theta}{3}+d^\a\,\sigma_{\a\b}\,d^\b+d^\a\,{\cal A}_\a\right)\equiv -E^2\;\bar{\cal H}_{\parallel}(E,d^\a)
\\ 
&&{\rm for}\;\;E_{\cal O}=1,\quad  \frac{dz(\l)}{d\l}= -(1+z(\l))^2\;\bar{\cal H}_{\parallel} \label{lz1}
\eea
where $\bar {\cal H}_{\parallel}$  represents the longitudinal Hubble expansion rate  in the direction of the  light ray.
In a LRS model we have the following decomposition along the radial space like vector $v^\a$ \cite{Umeh:2014ana} (see \ref{AppA})
\bea
{\cal A}_\a={\cal A}\,v_\a,\quad \sigma_{\a\b}=\Sigma\,v_{\a}\,v_{\b}
\eea
then, a further decomposition of the  spatial vector  $d^\a$ along   $v^\a$, gives
\be\label{grk}
d^a=\k\; v^a+\k^a,\quad v^a\,\k_a=0\quad {\rm and}\quad  d^ad_a=1\rightarrow \k_a\,\k^a=1-\k^2
\ee
with $\k\equiv \cos\xi$ the magnitude of the radial component or also the angle in between the radial direction $v^\a$ and the propagating photon $d^\a$ direction \cite{Koksbang:2022upf}.
Finally $\k^\a$ is a spatial vector living in a 2-d spherical sheet. Eq. (\ref{lz1}) translate in
\bea
 \frac{dz}{d\l}&=&- (1+z)^2\;\left(\frac{\theta}{3}+ \Sigma\,\cos\xi^2 + {\cal A}\,\cos\xi\right)\equiv -(1+z)^2\;
 \bar{\cal H}_{\parallel}(z,\xi)
\label{lz}
\eea
For a radial incoming photons we have $\cos\xi$=\,-1 and we  find the longitudinal Hubble expansion rate (\ref{hl}) \cite{Jimenez}
\be
{\cal H}_{\parallel}(z )\equiv\left(\frac{\theta}{3}+ \Sigma  - {\cal A} \right)
\ee
In general, the observer's  4-vector field $u^\m$ is geodesic, meaning it has zero acceleration.
However,   in our case, we have ${\cal A} \neq0$, which means we must  retain additional terms that are typically neglected.
Eq.(\ref{lz}) must be solved together with the rest of the geodesic eqs (\ref{eql}), which   result  in the following relations (\ref{hl})
\bea\label{geol}
\frac{dt}{dr}=-\frac{X}{F},\qquad \frac{dz}{dr}=(1+z)\,X\,{\cal H}_{\parallel}=(1+z)\,\frac{  X^{(1,0)}-  F^{(0,1)}}{F}
\eea

\section{Redshift Drift}\label{dri}
Another observable related to the redshift is his time variation. The so called Sandage-Loeb effect  characterises the change in redshift for a given source over time 
\cite{Sandage},\cite{Loeb}. 
Drift  signals arise from photons emitted  and received by the same  comoving objects at spatial coordinate ($r_{\cal E}=r$  and $r_{\cal O}=0$)  over time intervals.
The time variation of such observables leads to a series of time drift measurements, which can take different forms: position drift, angular drift, and redshift drift.
A direct measurement of these effects provides a model independent observation of the evolution of the Hubble flow in our Universe, in contrast to   conventional analyses  that relay on fitting formulas.\\
Let us consider two consecutive measurements of  the redshift by the observer ${\cal O}$
at two different 
times, $\tau_{\cal O}$ and $\tau_{\cal O}+\delta \tau_{\cal O}$ (we partially use the notations and the analysis of ref. \cite{Heinesen}).
From eqs (\ref{deck}) and (\ref{lz}), we can express the variation of the redshift with respect to the conformal time of the observer as 
\footnote{Note that $\frac{d \tau _{\cal E}}{d\tau _{\cal O}}=\frac{1}{1+z}=\frac{E_{\cal O}}{E_{\cal E}}$}
(see also (\ref{dir}))
\bea
 \dot z |_{\cal O}&=&\frac{dz}{d\tau _{\cal O}}=
 \frac{d }{d\tau _{\cal O}}\frac{E_{\cal E}}{E_{\cal O}}=
 \frac{E_{\cal E}}{E_{\cal O}} \left(\frac{d \tau _{\cal E}}{d\tau _{\cal O}} \,\frac{\dot E_{\cal E}}{E_{\cal E}} -
 \frac{ \dot E _{\cal O}}{ E _{\cal O}}\right)=
 -E_{\cal E}\,\int^{\l_{\cal O}}_{\l_{\cal E}}\,\left(\frac{\dot E}{E^2}\right)'\;d\l
\eea
We  now define and compute the following  directional derivatives of $E$ (\ref{dir}) ($E',\,\dot E,\,\hat E$)  \cite{Heinesen}
\bea
&&{\bf A}\equiv -\left(\frac{ E'}{E^3}\right)'=\left(\frac{ {\cal H}_{\parallel}}{E}\right)'=\Pi-{\bf I}={\cal H}_{\parallel}^2+\dot{\cal H}_{\parallel}-\hat {\cal H}_{\parallel}
\\
&& {\bf I}\equiv\left(\frac{\hat E}{E^2}\right)'=\dot{\cal H}_{\parallel}-\frac{\nabla_\a E}{E}\,{\cal L}_{\vec k}e^\a=
\hat {\cal H}_{\parallel}+{\cal H}_{\parallel}\,{\cal A}
\\
&& \Pi\equiv -\left(\frac{\dot E}{E^2}\right)'=\dot{\cal H}_{\parallel}-\frac{\nabla_\a E}{E}\,{\cal L}_{\vec k}u^\a=
 {\cal H}_{\parallel}^2+\dot{\cal H}_{\parallel}+ {\cal H}_{\parallel}\;{\cal A} 
\eea
here, ${\cal L}_{\vec k} $ is the Lie derivative operator along $k^\a$.
Using eq. (\ref{dirf}) for $f=E$, we obtain the following relationship
\bea\label{dirE}
E'=E\,(\dot E-\hat E)\;\to\; {\bf A}=-\left(\frac{ \dot E-\hat E}{E^2}\right)'\;\to\;
\Pi={\bf A}+{\bf I}
\eea
After integrating the above expressions, we find the following relationships  
\bea
&&E_{\cal E}\,\int^{\l_{\cal O}}_{\l_{\cal E}}{\bf A}\,d\l=
E_{\cal E}\;
\left.\frac{{\cal H}_\parallel}{E}\right|^{\l_{\cal O}}_{\l_{\cal E}}
=(1+z)\,{\cal H}_0-{\cal H}_\parallel(z)
\\
&&
E_{\cal E}\,\int^{\l_{\cal O}}_{\l_{\cal E}}\,{\bf I}\,\,d\l=
E_{\cal E}\,\int^{\l_{\cal O}}_{\l_{\cal E}}\,\left(\hat {\cal H}_{\parallel}+{\cal H}_{\parallel}\,{\cal A}\right)\,d\l=
E_{\cal O}\,(1+z)\,\int^{z}_{0}\,\frac{\hat {\cal H}_{\parallel}+{\cal H}_{\parallel}\,{\cal A}}{(1+z')^2\,{\cal H}_{\parallel}}\,dz'
\eea
and using the fact that
\bea
\frac{\left(\hat {\cal H}_{\parallel}+{\cal H}_{\parallel}\,{\cal A}\right)}{{\cal H}_\parallel}=
\frac{\hat {\cal H}_{\parallel}} { {\cal H}_{\parallel}}+\frac{\hat F} {F}= \hat \log({\cal H}_{\parallel}\,F)=\frac{\partial_r \log({\cal H}_\parallel\,F)}{X}
\eea
we obtain our final result (with $E_{\cal O}=1$) 
\bea\label{tzL}
\!\!\!\!\!\!\! \dot z|_{\cal O}&=&
(1+z)\,{\cal H}_0-{\cal H}_{\parallel}(z)+(1+z)\,\int^{\l_{\cal O}}_{\l_{\cal E}}\,{\bf I}\,d\l=
\\&&\nonumber
(1+z)\,{\cal H}_0- {\cal H}_{\parallel}(z)+
(1+z)\,\int^{z}_{0}\,dz'\,\frac{1}{(1+z')^2}\left.\frac{\partial_{r}\log ( {\cal H}_{\parallel}\,F)}{X}\right|_{r(z'),t(z')} =
\\&&
 (1+z)\,{\cal H}_0-\left.\frac{ X^{(1,0)}- F^{(0,1)}}{X\,F}\right|_{r(z),t(z)} \!\!\!\!\!\!\!+
(1+z)\,\int^{z}_{0}\,dz'\,\frac{1}{(1+z')^2}\left.\frac{\partial_{r}\log ( \frac{X^{(1,0)}-F^{(0,1)}}{X})}{X}\right|_{r(z'), 
t(z')} \nonumber
\eea
 This agrees   with the results obtained for    FRW and LTB spacetimes
\begin{itemize}
\item In FRW with $F=1,\,X=a(t),\,{\cal H}_{\parallel}=\frac{\partial_t a(t)}{a(t)}$ 
(from  eq. (\ref{HCDM}) and (\ref{arfrw}) we have ${\cal H}_{\parallel}(z)=H(a_0(z))\equiv H(z)$)   (see eq.(\ref{HCDM}) for the Hubble function in $\Lambda$CDM model)  we get \cite{Sandage}, \cite{Lobo:2022ubx}
\bea\label{tzfrw}
\dot z|_{\cal O}^{FRW}=(1+z)\,{H}_0-{H}(z)
\eea
 \item In LTB with $F=1$ and (\ref{ltbx}), ${\cal H}_{\parallel}=\frac{ X^{(1,0)}(t,r)}{X(t,r)}=\frac{ R^{(1,1)}(t,r)}{R^{(0,1)}(t,r)}$   we obtain \cite{Yoo:2011ibt}, \cite{Mishra:2014vga},\cite{ChirinosIsidro:2016vah}, \cite{Koksbang:2022upf}
\bea\label{tzltb}
\!\!\!\!\!\!\!\!\!\!\!\!\!\!
\dot z|_{\cal O}^{LTB}&=&(1+z)\,{\cal H}_0-{\cal H}_{\parallel}(z)
+(1+z)\,\int^{z}_{0}\,dz'\,\frac{1}{(1+z')^2}\left.\frac{\partial_{r}\log ( {\cal H}_{\parallel})}{X}\right|_{r(z'),t(z')}= \\&&(1+z)\,{\cal H}_0-{\cal H}_{\parallel}(z)
+(1+z)\,\int_{t(z)}^{t_0}\,dt'\,\frac{1}{(1+z(t',r(t')))}\left.\frac{\partial_{r}   {\cal H}_{\parallel} }{X}\right|_{t',r(t')}
\eea
 that corresponds exactly to eq. (25) of ref \cite{Codur:2021wrt}.
 
\end{itemize}
As discussed also in ref. \cite{Koksbang:2022upf} the simplifying local picture of a redshift drift $\dot z$ formula dominated by the first two terms to the right of eq. (\ref{tzL})  is not a good approximation.  

\section{Solutions near the center}\label{near}
In this section we use the cosmographic approach that consist in a Taylor expansion in redshift of the main cosmological observables that can be combined with a model independent interpretation of the data analysis \cite{Partovi:1982cg}.
 The solution of the eqs of motion  (\ref{eqF}, \ref{eqY}, \ref{eqX}) around the center at $r=0$ requires an expansion of the background space equation of the following form (where $\ar={\cal H}_0\,r$)
\bea
&&\Lambda(r)={\cal H}_0^2\,\left(\l_0+ \frac{\l_2}{2\,!}\,\ar^2+\frac{\l_3}{3\,!}\,\ar^3+\ldots\right),
\\&&\nonumber
{\bf m}(r)=\frac{1}{{\cal H}_0}\,\left(\frac{{\bf m}_3}{3\,!}\,\ar^3+\frac{{\bf m}_4}{4\,!}\,\ar^4+\ldots\right) ,
\\&&
{\bf E}(r)=\frac{{\bf e}_2}{2\,!}\,\ar^2+\frac{{\bf e}_3}{3\,!}\,\ar^3+\ldots ,\nonumber
\eea
where all the $\l_i,,{\bf m}_i,\,{\bf e}_i$ parameters are dimensionless.
The expanded solutions for the form factors of the metric result:
\bea
&&F=1+\frac{3}{4} \,a^3 \,\hat{r}^2\, \Omega _{\chi }+a^2 \,\hat{r}^3 \,\left(a \,\text{$\Delta $f}+R_2(a) \,\Omega _{\chi }\right)+\ldots
   \\
 &&  X=a+ \hat{r} \,R_2(a)+\frac{\hat{r}^2}{2} \, \left(R_3(a)-a^4 \, \Omega _{\chi } -a \,\Omega _{\kappa
   }\right)+\frac{\hat{r}^3}{6}\,   X_3(a)+\ldots
   \\
  && R=a \,r\,\left(1+\frac{\ar}{2}\, R_2(a)+\frac{\ar^2}{6} \,  R_3(a)\right)+\ldots
\eea
with $\text{$\Delta $f}\equiv\frac{6\, \l _3-\mathbf{m}_4\, \Omega _{\chi }}{108\, \Omega _m}$.
At   leading order ${\cal O}(r^0)$, we define
\bea
\l_0\equiv 3\,  \Omega_\Lambda,\;\;
\l_2\equiv \frac{9}{2}\,  \Omega_m\,\Omega_\chi,\;\;
{\bf m}_3\equiv 18\,  \Omega_m,\;\;
{\bf e}_2\equiv 2\,  \Omega_\k,\;\;
\eea
so that   we can write the {\it local Hubble flow} as \footnote{Such Hubble function is related
 to the Hubble flow around the center of the inhomogeneity (at $r=0$) and cannot be naively compared with the $\Lambda$CDM Hubble flow eq. (\ref{HCDM}).}
\bea
{\cal H}(a)\equiv\frac{a'(t)}{a(t)}=\frac{{\cal H}_0}{a^{3/2}}\;\sqrt{\Omega_m+a\,\Omega_\kappa+a^3\,\Omega_\Lambda+a^4\,\Omega_\chi}
\eea
We can begin by giving   physical meaning to some of the  parameters.
 The parameter $\Omega_\k$, which is proportional to the spatial curvature density,
 provides information about the local structure of the ${\bf E}$ form  factor.
 The matter density, proportional to $\Omega_m$, originates from ${\bf m}$. 
 The DE density, proportional to $\Omega_\Lambda$,  arises from ${\l}$,
  while an anomalous energy density parameter, $\Omega_\chi$, results
   from the simultaneous presence of  $\l$ and ${\bf m}$. This energy contribution has a very unusual equation of state, with $w=-4/3$.
 \\
 At second order the leading time dependent form factor $R_2(a)$ satisfy the differential eq
\bea 
 R_2'(a)&+&\left(\frac{1}{2} \left(\frac{\Omega _m}{a^4}-\frac{2\, \Omega _{\Lambda }}{a}-3\,  \Omega _{\chi }\right)\,R_2(a)-\frac{\mathbf{e}_3}{6\,  a^2}-\frac{\mathbf{m}_4}{72\,  a^3}-2 \, a \, \text{$\Delta $f}\right)\frac{  {\cal H}_0^2}{\mathcal{H}(a)^2}=0
\eea
with the   boundaries (at present time $a=1$) $R_2(1) =0$. 
We do not have to explicitly solve this equation but only evaluate
\bea
 R_2'(1)&=& 2 \, \text{$\Delta $f}+\frac{\mathbf{m}_4}{72}+\frac{\mathbf{e}_3}{6} \label{eq1r}
 \\ \nonumber
 R_2''(1)&=& \text{$\Delta $f}  \, \left(2  \, \Omega _{\kappa }+3 \,  \Omega _m-\Omega _{\chi }+4\right)+
 (12\,\mathbf{e}_3  +\mathbf{m}_4) \,  \left(\frac{\Omega _{\kappa
   }}{72}+\frac{\Omega _m}{48}-\frac{\Omega _{\chi }}{144}-\frac{1}{36}\right)+\frac{\mathbf{e}_3}{6}
\eea
that are useful for the next eqs.
At order three, the next correction $R_3(a)$ always evolves with a first order   equation  (with boundary $R_3(1)=0$) but   we need only  his first derivative value at present time
\bea\nonumber
R_3'(1)&=&
-3 \,\text{$\Delta $f}^2-\frac{\mathbf{e}_3  \,\mathbf{m}_4}{288}-\frac{\mathbf{e}_3^2}{48}+\frac{\mathbf{e}_4}{8}+\frac{\l_4}{9 \, \Omega
   _m}-\text{$\Delta $f}  \,\left(\mathbf{m}_4 \, \left( \frac{1}{9 \Omega _m}+\frac{1}{24}\right)+\frac{\mathbf{e}_3}{2}\right)+
  \\ &&\Omega
   _{\chi } \left(3\, \Omega _{\kappa }+\frac{9 \, \Omega _m}{4}+\frac{9}{2}\right)+\mathbf{m}_5 \, \left(\frac{1}{120}-\frac{\Omega _{\chi
   }}{72  \,\Omega _m}\right)-\frac{\mathbf{m}_4^2}{6912}+\frac{3  \,\Omega _{\chi }^2}{4}
\eea
The longitudinal and the transverse Hubble rates at first order are given by
\bea
&&\label{hp}
{\cal H}_{\parallel}(a,\,\hat r)={\cal H}(a)-\hat{r} \left( \frac{3}{2} \, a^2  \,{\cal H}_0 \,\Omega _{\chi }+\frac{\mathcal{H}(a)  \,\left(R_2(a)-a \,
   R_2'(a)\right)}{a}\right)
\\
&& 
{\cal H}_{\perp}(a,\,\hat r)={\cal H}(a)\left(1-\hat{r} 
\frac{   \,\left(R_2(a)-a \,
   R_2'(a)\right)}{a}
\right)
\\
&&\label{htheta}
{\cal H}_{\theta}(a,\,\hat r)=\mathcal{H}(a)\,\left(1-\frac{2\,\hat r}{3}\,
\frac{\left(R_2(a)-a \,
   R_2'(a)\right)}{a}\right)
   \eea
   where we note that     always ${\cal H}_{\parallel}(a, \hat r)\leq {\cal H}_{\perp}(a, \hat r) $ for $\Omega_\chi> 0$.

     \subsection{ Observables along the light cone }
  
  The perturbative solutions for the light cone geodesic eqs (\ref{geo1}), for small $z$ values, result \cite{Partovi:1982cg},
  \cite{Heinesen}, \cite{Asta1}
   \bea
\label{lca}&&   a(z)= 1-z+z^2\, \left(\frac{1}{2} \left(R_2'(1) +2\right)-\frac{3\,
   \Omega _{\chi }}{4}\right)+{\cal O}(z^3)
   \\\label{lcr}&&
   r(z)=\frac{z}{{\cal H}_0}+\frac{z^2}{{\cal H}_0}\, \left(\frac{1}{4} \,\left(2 \,\Omega _{\Lambda
   }-\Omega _m-2\, R_2'(1)-2\right)+\frac{3\, \Omega _{\chi
   }}{2}\right) +{\cal O}(z^3)
   \eea
   Parametrising   the small $z$ series of the luminosity distance function, $d_L(z)\equiv(1+z)^2\,R(a(z),\,r(z))$, in the following way \cite{Partovi:1982cg}
\bea
 \label{dLz}
d_L(z)&\simeq&\frac{z}{{\cal H}_0}+\frac{z^2}{{\cal H}_0}\,\frac{\left(1-Q_0\right)}{2}-
\frac{z^3}{{\cal H}_0}\,\frac{\left(1-Q_0-3 \,Q_0^2+J_0\right)}{6}+{\cal O}(z^4)
\eea
 we get, at first order, the   Hubble constant ${\cal H}_0$ that we set to come from   local measurements \cite{Riess}:   ${\cal H}_0=73.52\pm 1.62\,Km/s/Mpc$.
  At second order we have   the $Q_0$ deceleration parameter and, at third order, the jerk parameter $J_0$. 
In order to show the effect of the bubble profile, we can factorise out the $\Lambda$CDM contributions ($q_{0}$ and $j_0$)\footnote{
In the $\Lambda$CDM model, the small $z$ expansion of the Hubble function and of the effective  equation of state can be written as 
 \bea
 H_{\Lambda CDM}(z)&\equiv &\,{\cal H}_0\,\left(1+z(1+q_0)+\frac{z^2}{2}(j_0-q_0^2) +{\cal O}(z^3)\right)
 \\
 w_{\Lambda CDM}(z)&\equiv &\,   \frac{(-1+2\,q_0)}{3}+\frac{2\,z}{3}(j_0-q_0-2\,q_0^2) +{\cal O}(z^2) 
 \eea
where
\be\label{qjL}
q_{0}=-1+\frac{3}{2}\,\Omega_m,\quad j_{0}=1,\; \;\;{\rm with}\;\;\;\Omega_\Lambda+\Omega_m=1
\ee} writing $Q_0\equiv q_{0} +\Delta q,\; J_0\equiv j_{0} +\Delta j$
where
\bea\label{q0}
&&\!\!\!\!\!\!\!\!\!\!\!\!\!\!\! \Delta q= \Omega _{\kappa }-2\, \Omega _{\chi }+R_2'(1) 
\\\label{j0}
&& \!\!\!\!\! \!\!\!\!\!\!\!\!\!\!\Delta j=\frac{17\,  \Omega _{\chi }}{2}-6\,   \Delta  f -2 \, \Omega _{\kappa }+R_2'(1)\,  \left(2 \, \Omega _{\kappa }+3\,  \Omega
   _m-\Omega _{\chi }-1\right)-2\,  R_2''(1)+R_3'(1) 
\eea
Then we can compute   the $z$ series of the Hubble functions ${\cal H}_{i}(z)  =H_{\Lambda CDM}(z)+\Delta{\cal H}_{i}(z)$ and also of the effective equations of state 
 ${ w}_{i}(z) \equiv w_{\Lambda CDM}(z)+\Delta w_{i}(z)$, $\;i=\parallel,\,\perp,\,\theta$ , see (\ref{wzi}).\footnote{The metric form factors, along the null geodesics, at order ${\cal O}(z^3)$ result
 \bea
 F=1+\frac{3}{4}\,\Omega_\chi\,z^2,\quad
 R=\frac{z}{{\cal H}_0}\,\left(1-\frac{3+q_0+\Delta q}{2}\,z\right),\quad X=1-z+\left(1-\frac{\Delta q}{2}-\frac{9\,\Omega_\chi}{4}\right)\,z^2
 \eea note that the parameter $\Omega_\k$ is present only inside the $\Delta q$ (\ref{q0}) parameter.
 Eqs (\ref{lca}, \ref{lcr}), using eq. (\ref{q0}), can be rewritten as
 \bea
 a(z)=1-z+\left(1+\frac{\text{$\Delta $q}}{2}-\frac{\Omega _{\kappa
   }}{2}+\frac{\Omega _{\chi }}{4} \right)\,z^2,\qquad
   r(z)=\frac{z}{{\cal H}_0}\,\left( 1-\left(\frac{\text{$\Delta $q}}{2}+\frac{3 \,\Omega _m}{4}\right)\,z\right)
 \eea}
 For the   longitudinal Hubble expansion rate (\ref{hl}) we get
 \bea\nonumber
   {\cal H}_{\parallel}(z) &=&{\cal H}_0\, \left(1+z\,\left(1+Q_0\right)
   +\frac{z^2}{2}\,\left( \Delta  q +J_0-Q_0^2\right)\right)
   \\&&\nonumber
   \Delta{\cal H}_{\parallel}(z)=\left(z\;\,\Delta q+\frac{z^2}{2}\;\left(\Delta j-\Delta q^2+\Delta q\,(1-2\,q_0)\right)
    \right)\,{\cal H}_0\\&&
   \Delta w_{\parallel}(z)=  \frac{2 \,  \Delta  q }{3}+\frac{2}{3} \, z\, 
   \left( \Delta  j -2\,   \Delta  q ^2-4\, 
    \Delta  q \,  q_0\right)\label{hwL0}
   \eea
   so that a measure of $\Delta{\cal H}_{\parallel}(z)$ and or $\Delta w_{\parallel}$, at leading order in $z$, corresponds to a measure of $\Delta q$.
 For the transversal functions (\ref{eqY}) we get
 \bea
&&\label{Hpez}\Delta{\cal H}_{\perp}(z)=\Delta{\cal H}_{\parallel}(z)+\left(\frac{3}{2}\,z\,\Omega_\chi+
   z^2 \;\left(3\,\Delta f-\frac{3}{4}\left(\Delta q-(5+\,q_0)\right)\,\Omega_\chi\right) \right)\,{\cal H}_0 
 \\&&
   \Delta w_{\perp}(z)=\Delta w_{\parallel}(z)+\Omega_\chi+z\,\left(
   4\,\Delta f-3\,(2+\Delta q+q_0)\,\Omega_\chi-\frac{3}{2}\,\Omega_\chi^2 
   \right)\eea
    where we see that  a measure of $\Delta{\cal H}_{\perp}(z)$ and/or $\Delta w_{\perp}$, at leading order in $z$,
    corresponds to a measure of $\Omega_\chi$.
   For the volume Hubble expansion rate (\ref{eqY}) we get
    \bea\label{Hthz}
   \Delta{\cal H}_{\theta}(z) &=&  
     {\cal H}_0\,z\,\left(\frac{1}{6} \left(2 \,\Omega _{\kappa }+5\, \Omega _{\chi
   }\right)-\frac{\text{$\Delta $q}}{3}\right)+
      {\cal H}_0\, z^2 \,\left( \frac{5 \,\text{$\Delta $f}}{3}+\frac{\text{$\Delta
   $q}^2}{6}+\frac{1}{36} \text{$\Delta $q} \left(-2\,
   \Omega _{\kappa }+\right.\right. 
   \\&& \nonumber
   \left.\left.+15 \,\Omega _m-17\, \Omega _{\chi
   }-20\right)  +
    \frac{1}{72} \left(-8 \, \Omega _{\kappa }^2-16 \,
   \Delta j+\Omega _{\kappa }  \,\left(-30 \, \Omega _m+20 \, \Omega
   _{\chi }+8\right)-\right.\right. \nonumber
   \\&& \nonumber
   \left.\left.\Omega _{\chi } \left(21 \, \Omega
   _m+8  \,\Omega _{\chi }+88\right)-8  \,R_2''(1)\right)
   \right) 
   \eea
   In this case a measure of $\Delta {\cal H}_\theta$ is in between $\Delta {\cal H}_{\parallel/\perp}$ being proportional both to $\Omega_{\chi,\kappa}$.
 \\The $z$  expansion of the redshift drift in the $\Lambda$CDM model gives
   \bea
   \dot z_{\Lambda CDM}&=&-{\cal H}_0\,z \,q_0+{\cal O}(z^2)\eea
   while in our model we get (the details are given in appendix \ref{driftz}) \cite{Partovi:1982cg}
   \bea\label{dz0}
   \dot z&=&\dot z_{\Lambda CDM}+\Delta \dot z,\quad
   \Delta \dot z=-{\cal H}_0\,z\left(  \Omega_\k+\frac{\Omega_\chi}{2}\right) +{\cal O}(z^2)
   \eea
   so that a measurement of  $\dot z$,  together with  ${\cal H}_{\perp/\parallel}$,
   gives a bound on  $\Omega_\k$.
   From (\ref{sa}) we can get the leading terms for shear and   acceleration 
   \bea\label{rel0}
  && \Sigma = -\frac{2}{3} \,{\cal H}_0\, z \,\left( \Delta  q -\Omega _{\kappa }+2\, \Omega _{\chi
   }\right),\quad
   \mathcal{A}= \,{\cal H}_0\,  z\, \left(- \Delta  q +\Omega _{\kappa }-\frac{\Omega _{\chi
   }}{2}\right)
   \eea
   where we see as  the "anomalous" densities $\Omega_\chi$ and $\Omega_\k$ contribute differently to such a geometrical quantities.
   From (\ref{Hthz}, \ref{Hpez},  \ref{dz0}) at order ${\cal O}(z)$ we find  a consistency relationship in between different observables
   \bea\label{Hr}
   \left(\Delta{\cal H}_\theta-\frac{2}{3} \,\Delta{\cal H}_\perp+\frac{1}{3}\Delta\dot z\right)_{{\cal O}(z)}=0
   \eea
   or equivalently 
   \bea\label{wr}
   \left(\Delta{w}_\theta-\frac{2}{3} \,\Delta{w}_\perp \right)_{z=0} +
   \left(\frac{2}{9\,{\cal H}_0}\Delta\dot z\right)_{{\cal O}(z)}=0
   \eea
   that, in a sense,   corresponds to a {\it necessary condition} for the model  
   \footnote{For cosmological models that are described by a finite number of parameters (as the $\Lambda$CDM model)  the  {\it necessary conditions} can be applied taking into account the $z$ series expansion of only one observable, the luminosity distance $D_L$  \cite{Celerier:1999hp}.  }.
   Experimental observations, in the future, for the equations (\ref{Hr}) or (\ref{wr}) will be an interesting verification activity for the model.
  
  \section{Perturbative solution around FRW}\label{pertf}
In eq. (\ref{frwL}) we have  the parameter space  FRW limit of the model. Here we perturb the various form factors of the metric and the space dependent boundary conditions around such a values. 
For the  metric we define the following perturbative structure
\bea\label{exp}
 \!\!\!\! \!\!\!\! F=1+F_1+\frac{F_2}{2}+...,\;
   X=a(t)\left(1+X_1+\frac{X_2}{2}+...\right), 
  \; 
   R=a(t)\,r\,\left(1+R_1+\frac{R_2}{2}+...\right)
     \eea
     While, for the    space dependent functions, we introduce different dimensionless perturbations characterised by an index   "1" as $f_1$ ($f=\l,\,{\bf m},\,{\bf n},\,{\bf e}$) 
          \bea\label{lmne}
      &&{\Lambda}(r)\equiv \Lambda_0 \,(1+{\lambda}_1(r))
     ,\qquad 
     {\bf m}(r)\equiv  m_0\,r^3\,(1+{\bf m}_1(r))
     \\\nonumber
    && {\bf n}(r)\equiv  m_0\,r^2\,(1+{\bf  n}_1(r))
    ,\quad 
      {\bf E}(r)\equiv   { \bf e}_1 (r)= 2\,{\bf m}_1(r)+\frac{2}{3}\,r\,{\bf m}_1'(r)-2\,{\bf n}_1(r)
     \eea
To shorten expressions we introduce the notation \footnote{For early  times we have the limit
        $h(a\ll1)=1-\frac{2\,\Omega_\Lambda\,a^3}{11\,\Omega_m}+\ldots,$ 
        while for late time
        $h(a \gg 1)= \frac{2\,\Gamma(2/3)\,\Gamma(11/6)}
   {\pi^{1/2}}\left(\frac{\Omega_m }{ \Omega_\Lambda}\right)^{1/3}\frac{1}{a}+\ldots$,  
    today we have $h(a=1)\simeq 0.78$. }
\bea
h(a)\equiv \, _2F_1\left(\frac{1}{6},1;\frac{2}{3};-\frac{\Omega _\Lambda\,a^3 }{ \Omega _{m}}\right)
    \eea
  The $a=a(t)$ parameter is   the scale factor that satisfies the FRW Hubble time evolution equation
\be\label{HCDM}
H(a)\equiv\frac{a'(t)}{a(t)}=H_0\;\frac{ \sqrt{a(t)^3 \,\Omega _{\Lambda }+\Omega _m}}{a(t)^{3/2}},\qquad 
\Omega _{\Lambda }+\Omega _m=1
\ee
The solutions of the eqs of motion (\ref{eqF}, \ref{eqY}, \ref{eqX}) at first order   result
\footnote{It is important to stress the boundary conditions used to solve eqs (\ref{eqF}, \ref{eqY}). 
We try to impose the matching with a FRW metric at very early time ($a\to0$):
$F_1(0,r)=0$ and also $R_1(0,r)=0$. 
The solution (\ref{fxr}) for $F_1\sim a^3$ is fine. The generic solution for 
$R_1$ in the $a\to 0$ limit is of the form $ f(r)\,a^{-\frac{3}{2}}+\frac{{\bf m}_1(r)}{3}+{\cal O}(a)$ so that we can cancel with the boundary conditions only the singular term. The remaining ${\cal O}(a^0)$ term 
 leave the following limit
$R_1(0,r)=\frac{{\bf m}_1(r)}{3}$ and $X_1(0,r)=\frac{{\bf m}_1(r)+r\,{\bf m}_1'(r)}{3}$ that, with a change of  radial variable,  we can set to zero: ${\bf m}_1(r)=0$. In practice we have choose a gauge  \cite{Marra:2011zp} where ${\bf m}(r)=m_0\,r^3$, with $m_0=3\,H_0^2\,\Omega_m$.}
\bea\label{fxr}
\!\!\!F&=&1+a ^3 \,\frac{\Omega _{\Lambda }\, \lambda_1 (r)}{\Omega
   _m}\\\label{xfr}
   \!\!\! X&=&a\,\left(1+
   \frac{a  \,\mathit{h}(a ) \,\left(r\, \mathbf{e}_1'(r)-\mathbf{e}_1(r)\right)}{5\,
   H_0^2 \,r^2\, \Omega _m}-\frac{\mathbf{e}_1(r)}{2}+
   \frac{a ^3 \,  \Omega _{\Lambda }\, \lambda_1 (r) }{3\,  \Omega _m}-
   \frac{ a  \,(\mathit{h}(a )-1)\, 
   \lambda_1 ''(r)}{3\, H_0^2\, \Omega _m}
    \right)\\\label{rfx}
  \!\!\! R&=&a\,r\,\left(1+\frac{a \, \mathit{h}(a )\, \mathbf{e}_1(r)}{5\, H_0^2\, r^2\, \Omega _m}-\frac{a \,
   (\mathit{h}(a )-1) \,\lambda_1 '(r)}{3 \,H_0^2\, r\, \Omega _m}+\frac{a ^3\, \Omega
   _{\Lambda } \,\lambda_1 (r)}{3 \,\Omega _m}+\frac{\mathit{m}_1(r)}{3} \right)
\eea
 Note that to have regular functions for $r\to 0$,  we need at least
  \be\label{rto0}
   \frac{\l_1'}{ r},\;\;  
    \frac{ {\bf e}_1}{r^2},\;\; \frac{ {\bf e}_1'}{r}\sim finite
     \ee
that requires a small $r$ expansion of the form
     \be
   \l_1\sim const+\# \,r^2,\;\;\;  
      {\bf e}_1\sim r^2, \;\;\;\;{\rm i.e.}\;\;\;\;
 \lambda_1'(0)=0,\quad \mathbf{e}_1(0)=\mathbf{e}_1'(0)=0
\ee
The function $\mathbf{m}_1(r) $ can be reabsorbed through  a radial redefinition (see footnote 3),  effectively allowing us to set $\mathbf{m}_1(r) =0$ everywhere in (\ref{lmne}) and (\ref{fxr}, \ref{xfr}, \ref{rfx}).
 Note that, with respect the previous analysis in \cite{Comelli:2023otq}, where only the CC was perturbed, we now have an additional   free form factor, ${\bf e}_1(r)$, which is proportional to the density perturbations in eqs (\ref{lmne}).
 \\
 At infinity, we impose the boundary conditions $\l_1(r),\,{\bf e}_1(r)\to 0$ ensuring that the Hubble function asymptotically approaches  the Planck value.
 The perturbativity of the system is guarantied as long as the following conditions hold: 
 \bea
 a^3\,\l_1(r),\;a\,\frac{{\l}_1'(r)}{H_0^2\,r },\;a\,\frac{{\bf e}_1(r)}{H_0^2\,r^2},\;a\,\frac{{\bf e}_1'(r)}{H_0^2\,r}<1
 \eea
 These conditions generally define a  closed region in the 2-d spacetime, in the variables $a-r$. However, since $a=1$ at present time, we can  safely  neglect   the time dependence in the above inequalities.
 \subsection{Redshift equations}\label{lux}
In this context the perturbative solutions of the    geodesic eqs (\ref{geol}) 
can be obtained making a
change of time variable from $t\to a(t)$ where
$dt=\frac{da}{a\;H(a)}$. In this way   each time derivative has to be substitute with
$
\partial_t f(t,r)=a\,H(a)\,\partial_a f(a,r)
$
and the system of redshift-geodesic eqs  (\ref{geol}) becomes
\be\label{geo1}
\frac{da}{dz}=-\frac{a\;H(a)}{(1+z) \;F\;{\cal H}_{\parallel}}
,\qquad
\frac{dz}{dr}=(1+z) \;X\;{\cal H}_{\parallel} , \qquad \frac{da}{dr}=-\frac{a\;H(a)\;X}{ F }
\ee
To treat  the above eqs, once we know the functional dependence of the metric (\ref{fxr}, \ref{xfr}, \ref{rfx}) and Hubble (\ref{hl}), we  perturbatively   set  $a(z)=a_0(z)+a_1(z)+ \dots$ and $\;r(z)=r_0(z)+r_1(z)+ \dots $ and  then we solve order by order.
At zero  order we get the FRW results
\be\label{arfrw}
\!\!\!\!\!\! \!
a_0(z)=\frac{1}{1+z},\;\;
r_0(z)=\frac{2}{H_0\,\Omega_m^{1/2}}\left(  _2F_1\left(\frac{1}{6},\frac{1}{2};\frac{7}{6};-\frac{\Omega _{\Lambda }}{\Omega
   _m}\right)-\frac{\, _2F_1\left(\frac{1}{6},\frac{1}{2};\frac{7}{6};-\frac{\Omega _{\Lambda
   }}{(z+1)^3 \Omega _m}\right)}{\sqrt{z+1}}
\right)
\ee
while at first order   we obtain the following  system of  first order differential eqs 
\bea\nonumber
 a_1'(z)&+&\frac{a_1(z)}{ 1+z}+ \frac{\left(5-3 \, h(a_0)\right)
   \left(\mathbf{e}_1(r_0)-r_0\, 
   \mathbf{e}_1'(r_0)\right)}{10\,  r_0^2\, 
   H(a_0)^2}+\frac{\Omega _{\Lambda } \, \left((1+z)\,  \lambda
   '(r_0)-\lambda (r_0)\, 
   H(a_0)\right)}{(1+z)^5\,  \Omega _m\, 
   H(a_0)}+\\
   &&\label{eqa1}
   \frac{\lambda ''(r_0)\, \left(-3 \, (1+z)^3\, 
   \left(h(a_0)-1\right) \, \Omega _m-2\, 
   \Omega _{\Lambda }\right)}{6 \, (1+z)^3\,  \Omega _m\, 
   H(a_0)^2}=0
\eea
\bea\nonumber
&&r_1'(z)-\frac{H_0^2\, (1+z)\, \left((1+z)^3\, \Omega _m-2\,  \Omega
   _{\Lambda }\right)}{2 \, H(a_0)^3}\,a_1(z)+
   \frac{\Omega _{\Lambda }\,  \left(\lambda
  (r_0) \, H(a_0)-3\,  (1+z)\,  \lambda_1
   '(r_0)\right)}{3 \, (1+z)^3\,  \Omega _m\, 
   H(a_0)^2}+\\
   &&\label{eqr1}
   \frac{\lambda_1 ''(r_0) \left((1+z)^3\, 
   \left(h(a_0)-1\right) \, \Omega _m-2\, 
   \left(h(a_0)-2\right) \, \Omega _{\Lambda
   }\right)}{6 \, (1+z) \, \Omega _m\,  H(a_0)^3}-
   \frac{\mathbf{e}_1(r_0)}{2 \,H(a_0)}
+
\\&&\nonumber
\frac{\left(\mathbf{e}_1(r_0)-r_0\, 
   \mathbf{e}_1'(r_0)\right)\,  \left((1+z)^3\, 
   \left(h(a_0)-5\right)\,  \Omega _m-2\, 
   h(a_0) \, \Omega _{\Lambda }\right)}{10\, 
   r_0^2\,  (1+z)\,  \Omega _m\,  H(a_0)^3}=0
   \eea
where    $a_0=a_0(z)$ and  $r_0=r_0(z)$  (see (\ref{arfrw})) and with initial conditions $a_1(0)=r_1(0)=0$.
   
\subsection{Hubble functions }\label{hubfrw}
\begin{figure}
\centering
\begin{minipage}{.5\textwidth}
  \centering
  \includegraphics[width=.9\linewidth]{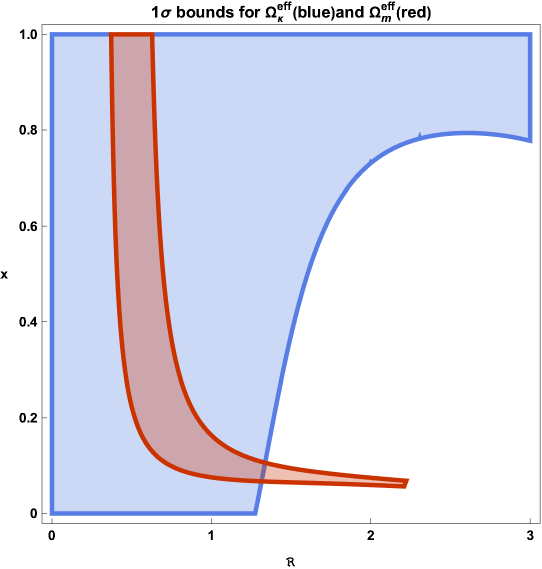}
\end{minipage}%
\begin{minipage}{.5\textwidth}
  \centering
  \includegraphics[width=.9\linewidth]{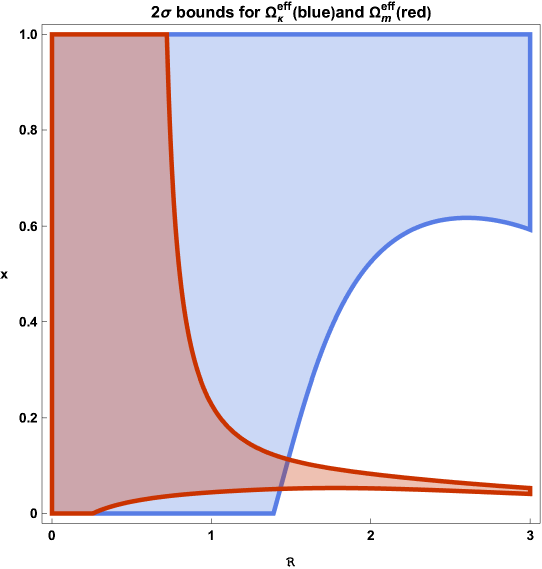}
\end{minipage}
\caption{We implement in the parameter space $(x-\Re)$ 
the constraints (\ref{planckdata}) for   $\Omega_m^{eff}$ (in red) and $\Omega_{\k,\parallel}^{eff}$ (in blue).
{  On the left} we have the $1\;\sigma$ bounds and {  on the right} the $2\;\sigma$ bounds. The overlap of the two figures will be the parameter space that we consider in the rest of the paper.}
          \label{figOMk}
\end{figure}

\begin{figure}
\centering
\begin{minipage}{.5\textwidth}
  \centering
  \includegraphics[width=.9\linewidth]{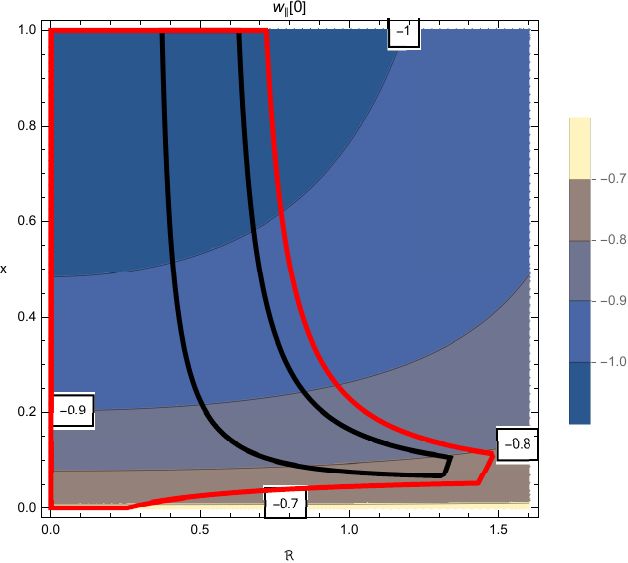}
\end{minipage}%
\begin{minipage}{.5\textwidth}
  \centering
  \includegraphics[width=.9\linewidth]{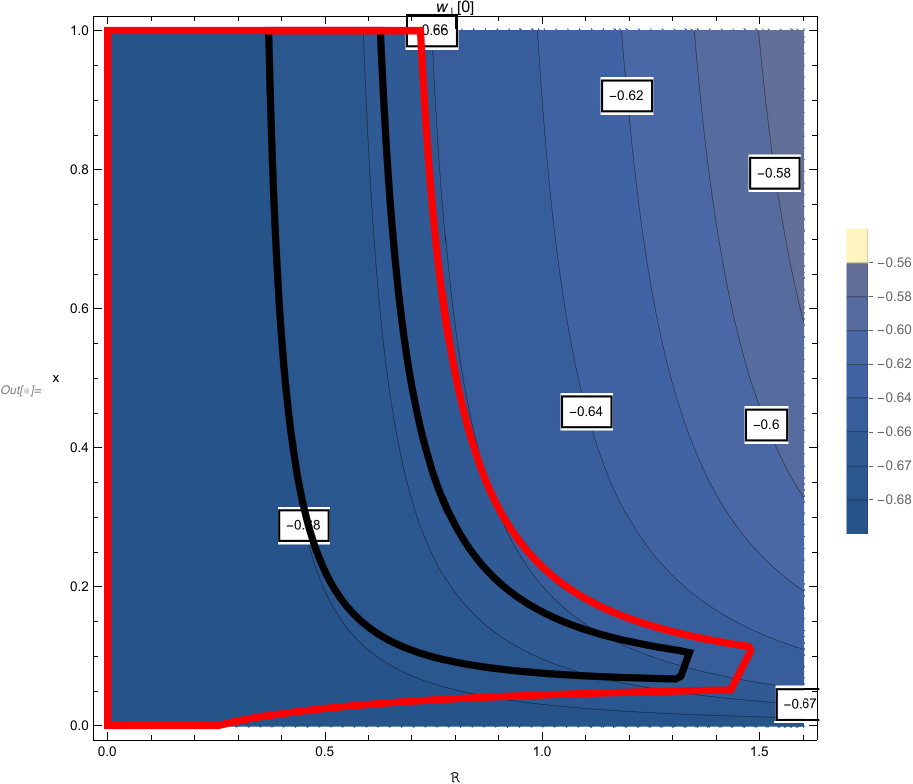}
\end{minipage}
\caption{On the left, we show   the contourplot of $w_{\parallel}(0)$ (\ref{eqwL0}) and 
on the right, the contourplot of $w_{\perp}(0)$ (\ref{eqw}). The black and red curves denote the $1-2\,\sigma$  regions from  fig.\ref{figOMk}.}
\label{figW0}
\end{figure}

Now that we have the perturbative solutions  of the Einstein equations and of the geodesic light cone, we can start  studying the  Hubble functions at first order 
\bea\label{ht}
\!\!\!\! {\cal H}_{\perp}(a,r)^2&=&H(a)^2+ \left(\frac{\mathit{h}(a)-1}{a^2}+\frac{2\,  a\,  \Omega _{\Lambda }}{3\,     \Omega
   _m}\right)\,\frac{\lambda_1 '(r)}{r}+\frac{(5-3\,  \mathit{h}(a))}{5 \, a^2 }\, 
  \frac{ \mathbf{e}_1(r)}{  r^2}\\\nonumber
\!\!\!\! {\cal H}_{\parallel}(a,r)^2&=&H(a)^2\,-
\frac{2 \, a^2\,  H(a)\,  \Omega _{\Lambda }\, }{\Omega _m}\, \lambda_1
   '(r)+
   \left(\frac{\mathit{h}(a)-1}{a^2}+\frac{2 \, a\, 
   \Omega _{\Lambda }}{3 \, \Omega _m}\right)\,\lambda_1 ''(r)-
   \\&&\label{hhl}
   \frac{(5-3\, \mathit{h}(a)) }{5 \, a^2  }\,\frac{\left(\mathbf{e}_1(r)-r\, 
   \mathbf{e}_1'(r)\right)}{ r^2}
   \\ 
\!\!\!\! {\cal H}_{\theta}(a,r)^2&=&H(a)^2 +
   \frac{
   \left(2 \, a^3 \, \Omega _{\Lambda }+3\,  (\mathit{h}(a)-1)
   \Omega _m\right)}{18 \, a^2\,  r \, \Omega _m}\,\left(2 \, \lambda _1'(r)+r\,  \lambda _1''(r)\right)+
  \\&&\nonumber \frac{(5-3 \, \mathit{h}(a)) }{30 \, a^2 \, r^2}\,\left(\mathbf{e}_1(r)+r\, 
   \mathbf{e}_1'(r)\right)
   \eea
Defining the Hubble deviations from the FRW evolution  as ${\cal H}_{\perp/\parallel}(a,r)=H(a)+\Delta{\cal H}_{\perp/\parallel}(a,r)$ we get, at present time $(a=1)$, the relationship
\bea
\Delta{\cal H}_{\parallel}(1,r)= \partial_r\,\left(r\,\Delta{\cal H}_{\perp }(1,r)\right)-\frac{\Omega_\Lambda}{ \Omega_m}\l_1'(r)
\eea
       It is important to set the feature of our model both around the center of symmetry than in the far past 
       along the null geodesic light path.
       \\
 At present time ($a=1$), the two Hubbles parameters coincide  at the center ($r=0$) of the inhomogeneity 
 (see also (\ref{rto0})) and defining  ${ \cal H}_0 \equiv{ \cal H}_{\perp/\parallel}(a=1,r=0)$, we get:
\be\label{lp}
\underbrace{{ \cal H}_0}_{\rm Local}\equiv \underbrace{H_0}_{\rm Planck}+\Delta H_0
= H_0+\frac{1}{H_0}\,
\underbrace{ \left(\frac{\mathit{h}(1)-1}{2}+
 \frac{ \Omega _{\Lambda }}{3\,   \Omega_m}\right)}_{\sim 0.667}\,\lambda_1''(0) +
 \frac{1}{H_0}\underbrace{\frac{(5-3\,  \mathit{h}(1))}{20 }}_{\sim 0.133}\, 
    \mathbf{e}_1''(0) 
\ee
 for $H_0$ we take the Planck data \cite{Planck} while for ${ \cal H}_0$ the local measurement \cite{Riess}
\be
{ \cal H}_0=73.52\pm 1.62\,\#,\quad H_0=H_0^{CMB}=67.4\pm 0.5\,\#,\quad 
\Delta H_0=6.12\pm1.7\,\#,\label{plD}
\ee
where $
\#= \frac{Km}{sec  \,Mpc}$, 
for the rest of the parameters we fix $\Omega_m=0.3$ and $\Omega_\Lambda=0.7$.\\
We can   replace the initial values $(\lambda_1''(0),\,\mathbf{e}_1''(0))$ in (\ref{lp}) with two more physical parameters $ (x,\,\Delta\l)$ :
\bea\label{parl}
&& \lambda_1''(0)\equiv H_0^2\,x\;\Delta\l ,\qquad  \mathbf{e}_1''(0)\equiv H_0^2\,(1-x)\;\Delta\l ,\\
 \label{dl}
&&\Delta\l=\left(\frac{60\,\Omega_m}{20\,x+(13\,x-3)\,\Omega_m\,(3\,h(1)-5)}\,
\frac{\Delta H_0}{H_0}\right)\simeq
\left(\frac{1.87}{0.25+x}\,
\frac{\Delta H_0}{H_0}\right)
\eea
in this way with $\Delta\l$ we define the amplitude of the perturbations while 
with the parameter $x$ we weigh the different components:  for $x\to 1$  are present only DE perturbations while for $x\to 0$ only the DM component perturbations 
($x=\l_1''(0)/(H_0^2\,\Delta\l)$ and $\Delta\l=(\mathbf{e}_1''(0)+\l_1''(0))/H_0^2$).
To have $\Delta H_0>0$, we naively require $ \mathbf{e}_1''(0)>0$ and  $\l_1''(0)>0$ i.e.,
 $\Delta\l>0$ and $0<x<1$.
Furthermore, we observe that for fixed $\Delta H_0$, the corresponding $\Delta \l$ for a pure DM ($x=0$) component is five times larger that for a  pure DE component ($x=1$). In this  sense,
 we can say that DE is "{\it heavier}" than DM.
Now that we have determined  the size of the present-time corrections, we can proceed to analyse the full functional form of the unknown functions.
\\ 
We choose the simplest and  most economical functional form for the full $r$-dependent  contributions of DE ($\l_1(r)$) and DM (${\bf e}_1(r)$), 
involving only one additional   free parameter, $\Delta$, which is related to the tail of the  corrections.
 To work  with dimensionless quantities, we    define  $\Delta\equiv\frac{ {  \Re}}{H_0}$ \footnote{  $\Re$ is a dimensionless parameter  that represents the size of the bubbles for ${\cal H}(r), \, {\bf e}_1(r)$ and $\l(r)$ in terms of $H_0$.}.
The functional structure is  (that matches the expansion (\ref{parl}))
\be\label{le1}
\lambda_1(r)=-\frac{\Delta \lambda}{2}\,H_0^2\,\Delta^2\;x\;e^{-\frac{r^2}{\Delta^2}},\quad
{\bf e}_1(r)=\frac{\Delta \lambda}{2}\,H_0^2\,r^2\,(1-x)\,e^{-\frac{r^2}{\Delta^2}},\;\;\;\Re=\Delta\,H_0
\ee
where you can use (\ref{dl}) for $\Delta \lambda$.
 Note that while $\l_1(r)$ is negative defined, ${\bf e}_1(r)$ is a positive function. 
 With this choice the corrections to transverse Hubble is simply
\be\label{leh1}
\Delta {\cal H}_{\perp}(1,r)=\Delta {  H}_0\;e^{-\frac{r^2}{\Delta^2}}
\ee
 In general, since $\Delta\l$ is fixed by eq. (\ref{dl}),  we are left with
  two free parameters to consider: 
  $x$, the fraction of DE to DM ($x\to 0$ corresponding  to pure ${\bf e}_1$ and
  $x\to 1$  corresponding to pure $\l_1$), and $\Re$,   the dimensionless  Hubble number that characterises the size of the spherical bubble. 
   This exponential structure simplifies  the calculations, as each radial derivative is proportional to the function itself. 
 Using these functions, we can evaluate the corrected form of the metric elements and the geodesic eqs.
  The Hubble functions, evaluated along the light of sight, ${\cal H} (z)\equiv {\cal H}(a(z),\,r(z))$,
 serve as the first interesting probe for the early-time cosmology.
 \begin{figure}
\centering
\begin{minipage}{.5\textwidth}
  \centering
  \includegraphics[width=.9\linewidth]{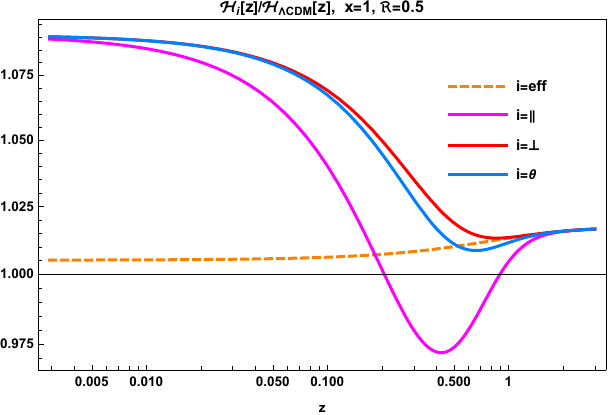}
\end{minipage}%
\begin{minipage}{.5\textwidth}
  \centering
  \includegraphics[width=.9\linewidth]{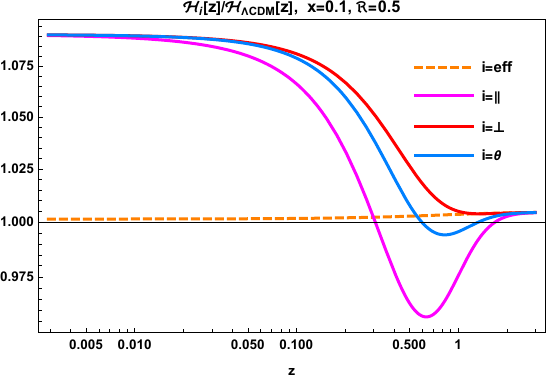}
\end{minipage}
\caption{Here the $z$ evolution of the ratios   between the asymptotic value of 
${\cal H}_{\perp}(z\gg1)$ of eq. (\ref{Heff}) that we call  ${\cal H}_{eff}$, the longitudinal 
${\cal H}_{\parallel}(z)$  of eq. (\ref{hl}), the transverse 
${\cal H}_{\perp}(z)$ of eq. (\ref{eqY}) and the  local 
${\cal H}_{\theta}(z)$  of eq. (\ref{Htheta}) over the ${\cal H}_{\Lambda CDM}(z)$ of (\ref{HCDM}), for fixed $\Re=0.5$  with   $x=1$ (only DE) to the left and $x=0.1$ (mainly DM) to the right.}
          \label{figH}
\end{figure}
    To probe the matching with the FRW early-time dynamics \cite{Biswas:2010xm}, we need to
    expand our formulas in the large $z$ limit.
 In the asymptotic limit $z\gg1$ (i.e. $a_0(z)\to 0$), we find that  $r_0(z)$ approaches  a finite value, which corresponds to the event horizon of the FRW  model, $r_0(z)\to r_H \equiv\frac{2 \, }{H_0\, \sqrt{\Omega _m}}\;_2F_1\left(\frac{1}{6},\frac{1}{2};\frac{7}{6};-\frac{\Omega _{\Lambda
   }}{\Omega _m}\right)\sim\frac{3.3}{H_0}$.
 For the transverse expansion rate  (\ref{eqY}) we can then  express the asymptotic expansion in terms of the improved effective densities
   \bea\label{Heff}
   {\cal H}_{eff}(z)^2\equiv{\cal H}_\perp^2(z\gg1)&\simeq& H_0^2\left(\frac{\Omega^{eff}_m}{a_0(z)^3}+\frac{\Omega^{eff}_{\k,\perp}}{a_0(z)^2}+ \Omega^{eff}_\Lambda \right)+{\cal O}(a_0(z))
   \\ 
   &&\hspace{-4.cm}  \Omega^{eff}_m=\Omega_m\,\left(1-3\,\frac{a_1(z)}{a_0(z)}\right)_{z\gg 1},\quad
   \Omega^{eff}_{\k,\perp}=\frac{2}{5}\,\frac{\,{\bf e}_1(r_H)}{r_H^2\,H_0^2},\quad \Omega^{eff}_\Lambda=\Omega_\Lambda\label{Op}
   \eea
   where  the ratio $\left(\frac{a_1(z)}{a_0(z)}\right)_{z\gg 1}$ in obtained integrating eq. (\ref{eqa1}) and it results a finite number.
   Note that for the other Hubble rates (\ref{hl}, \ref{Htheta}), we found some differences
    \bea\nonumber
   &&{\cal H}_\parallel^2 \sim  H_0^2\left(\frac{\Omega^{eff}_m}{a_0(z)^3}+
   \frac{\Omega^{eff}_{\k,\parallel}}{a_0(z)^2}+ \Omega^{eff}_\Lambda\right)+{\cal O}(a_0(z))
\\&&\nonumber
      {\cal H}_\theta^2\sim H_0^2\left(\frac{\Omega^{eff}_m}{a_0(z)^3}+
   \frac{\Omega^{eff}_{\k,\theta}}{a_0(z)^2}+ \Omega^{eff}_\Lambda\right)+{\cal O}(a_0(z))\\
  && \Omega^{eff}_{\k,\parallel}
  =\left(1-2\frac{r_H^2}{\Delta^2}\right)\,\Omega^{eff}_{\k,\perp},\qquad 
   \Omega^{eff}_{\k,\theta}=  
   \left(1-\frac{2}{3}\,\frac{r_H^2}{\Delta^2}\right)\,\Omega^{eff}_{\k,\perp}\label{OL}
   \eea
    where we see that while   $\Omega^{eff}_{\k,\perp}$   is positive defined (\ref{le1}), 
     $\Omega^{eff}_{\k,\parallel} $ and  $\Omega^{eff}_{\k,\theta} $ are positive  only for small   ratios $\frac{r_H^2}{\Delta^2}\sim\frac{10}{\Re^2} $ i.e. for   bubbles with $\Re\geq 0.07$ and $\Re\geq0.04$.
   The  $1\,\sigma$ constraints,  from the combination of Planck, lensing and BAO data \cite{Planck}, 
   for the spatial curvature and the matter density are given by
   \be\label{planckdata}
   \Omega_\k=0.0007\pm0.0019,\qquad \Omega_m=0.3081\pm 0.0065
   \ee
   In the fig.\ref{figOMk}, we apply   the constraints on $\Omega^{eff}_{\k,i}$ and $\Omega^{eff}_{m}$, which    determine the  allowed $x-\Re$ parameter space  at 1-2 $\sigma$ level. 
   The  strongest constraint, among the three components $\Omega^{eff}_{\k,i}$, 
   comes from $\Omega^{eff}_{\k,\parallel}$ and at 1 $\sigma$ level, the left fig.\ref{figOMk}, is pushing for values of  $\Re\sim 0.4-0.6$     for large $x$ or for small $x$ values  (always $x\geq 0.06$) it is possible also $\Re\sim 2.3$. In any case the 2 $\sigma$ level bounds are much more approximate, see  the right fig.\ref{figOMk}.
   This initial constraint on the parameter space arises from matching with the early-time cosmology.
   For the late time cosmological parameters ,  we present   figures for various quantities:  $w_{\parallel,\perp}(0)$ (see fig.\ref{figW0}), the present-time deceleration parameter  $Q_0$ (see fig.\ref{figQJ0} on the left), the jerk parameter   $J_0$ (see fig.\ref{figQJ0} on the right), and finally,   the first coefficient of the redshift drift $\frac{\dot z(0)}{z}$ (see fig.\ref{figDz0}).
 %
 \subsubsection{Effective equation of state}\label{wfrw}
 An interesting function that characterises the $z$-derivative of the Hubble functions  ${\cal H} (z) $ are the effective equation of state so defined
 \be\label{wzi}
 w_i(z)=-1+\frac{2}{3}\,\frac{d\,{\cal H}_i(z)}{d\,\log(1+z)},\qquad i=\parallel,\perp,\theta
 \ee
 Their general value at $z=0$ is given by
\bea\label{eqw}
w_{\perp}(0)&=&-\Omega_{\Lambda }+
   \frac{2 \,  \Delta \mathcal{H}'(0)}{3  \, H_0^2}+\frac{ \Delta  H_0}{H_0}\,  \left(\frac{10}{3\,  (3\,  \mathit{h}(1)-5)}+2\,  \Omega _{\Lambda
   }\right)-\lambda (0) \, \Omega _{\Lambda }
 \\\label{eqwL0}
w_{\parallel}(0)&=&w_{\perp}(0)+\frac{2\,\lambda (0)}{\Re^2}\, \left(1+\frac{20-14\,  \mathit{h}(1)}{3\,  (3\, 
   \mathit{h}(1)-5) \, \Omega _m}\right)
\eea
once we choose the functional form (\ref{le1},\,\ref{leh1}) we have ($w_{\Lambda CDM}(0)=- \Omega_\Lambda=-0.7$)
\be
w_{\perp}(0)=w_{\Lambda CDM}(0)+\left(0.13+\frac{0.06\,x\,\Re^2}{0.25+x}\right),\qquad
w_{\parallel}(0)\simeq w_{\perp}(0)-\frac{0.475\,x}{0.25+x}
\ee
that implies  $ \left.w_{\parallel}(0)\right|_{x=0}<w_{\perp}(0)<\left.w_{\parallel}(0)+0.38\right|_{x=1}$. Further the effective eqs of state   have a minimum, one  for $w^{min}_{\perp}(0) =-0.69$ when $\Re\to 0$ or $x\to 0$ and one for
$w^{min}_{\parallel}(0) =-1.07$ when $x\to1,\,\Re\to0$ (note the fact that $w_{\parallel}$ can be less than -1 for a period \cite{carroll}).

    \begin{figure}
\centering
\begin{minipage}{.5\textwidth}
  \centering
  \includegraphics[width=.9\linewidth]{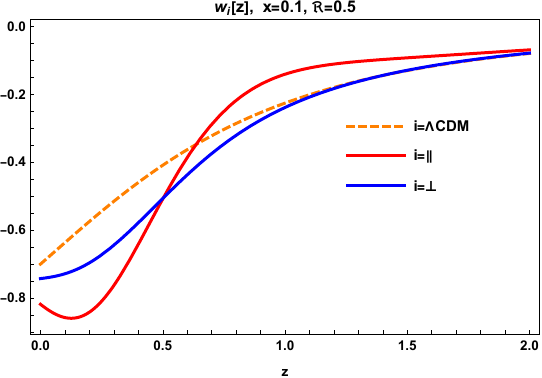}
\end{minipage}%
\begin{minipage}{.5\textwidth}
  \centering
  \includegraphics[width=.9\linewidth]{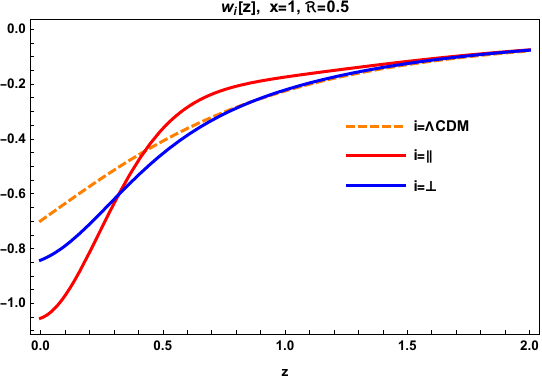}
\end{minipage}
\caption{The $z$ evolution of various  effective equation of states $w_{i}(z)$ (\ref{wzi}):   for $i=$$\Lambda$CDM , $i=\parallel$ longitudinal and $i=\perp$ transverse.
 On the left  for $x=0.1,\,\Re=0.5$.
  On the right  with $x=1,\,\Re=0.5$.
}
\label{figWz}
\end{figure}

    \subsection{Deceleration and Jerk}\label{qjfrw}
\begin{figure}
\centering
\begin{minipage}{.5\textwidth}
  \centering
  \includegraphics[width=.9\linewidth]{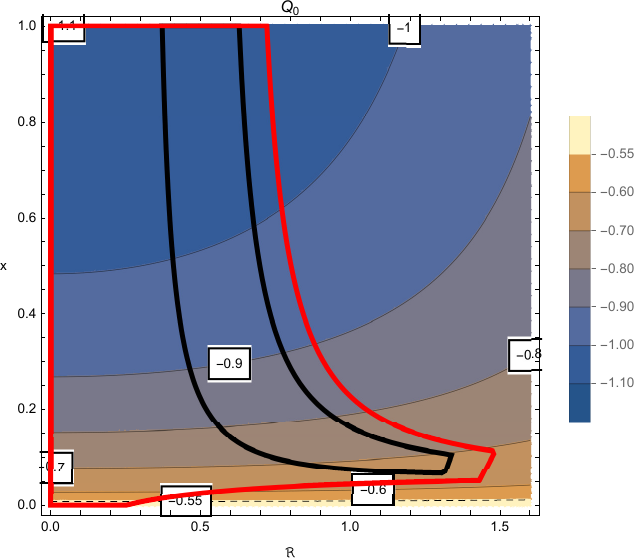}
\end{minipage}%
\begin{minipage}{.5\textwidth}
  \centering
  \includegraphics[width=.9\linewidth]{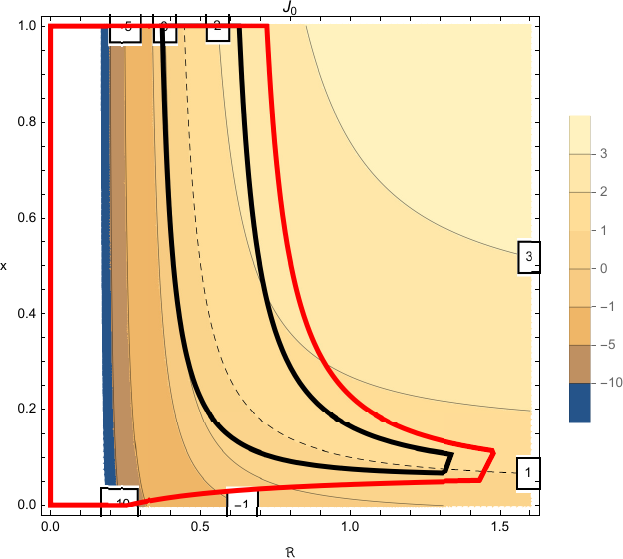}
\end{minipage}
\caption{On the left we plot the value of the deceleration $Q_0$ (\ref{dec}). 
On the right, the value of the jerk $J_0$ (\ref{jer}).
In black and red,  the $1-2\,\sigma$ region of fig.\ref{figOMk}. The dashed lines are the corresponding $\Lambda$CDM values.}
\label{figQJ0}
\end{figure}

    Analogously to  the effective eq. of state (\ref{wzi}) we can define the deceleration and jerk $z-$ dependent parameters as
    \bea\label{qzi}
    &&q_i(z)=-1+ \frac{d\,{\cal H}_i}{d\log(1+z)},\quad i=\parallel,\,\perp,\,\theta
    \\ && j_i(z)=(1+2\,q_i)\,q_i+(1+z)\;\frac{d q_i}{dz}
    \eea
and then (see also (\ref{dLz})) we can evaluate the deceleration $Q_0=q_{\parallel}(0)$ and the jerk parameter $J_0=j_\parallel (0)$ at present time to be (see (\ref{qjL}) for the $\Lambda$CDM values)
     \bea
    && \label{dec}
   Q_0= q_\Lambda+ \left(  -0.69+\frac{0.18+0.09\,x\,\Re^2}{x+0.25}\right)
     \\&& \label{jer}
      J_0=  j_\Lambda+ \left(3.61 -\frac{1.07 }{ x+0.25} -\frac{0.54}{\Re^2}\right)
    \eea
where we see that for $Q_0$ we have a minimum $Q_0^{min}=-1.24$ for $\Re\to 0$ and $x\to 1$,
 while for $J_0$ we note the singularity $1/\Re^2$ that set strong bounds for small bubbles.
In the literature the limits on such a parameters are strongly dependent from their $z$-functional form    \cite{AlMamon:2018uby}. 
 In ref \cite{Camarena:2019moy}  (where only supernovae in the redshift range $0.023 \leq z \leq 0.15$ are used for the fit) it is reported  the following limit $Q_0=-1.08 \pm 0.29$ that, looking to fig.\ref{figQJ0},  indicates a bubble dominated by a DE component with large $x$ values.  
      \begin{figure}
\centering
\begin{minipage}{.5\textwidth}
  \centering
  \includegraphics[width=.9\linewidth]{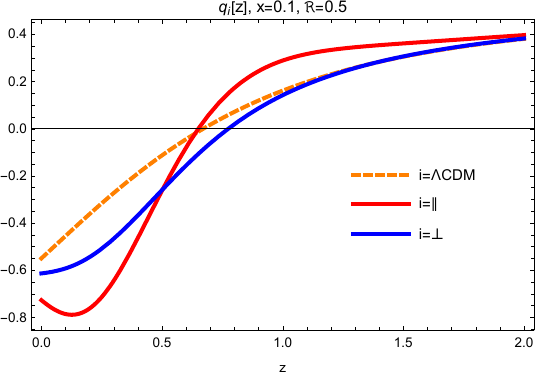}
\end{minipage}%
\begin{minipage}{.5\textwidth}
  \centering
  \includegraphics[width=.9\linewidth]{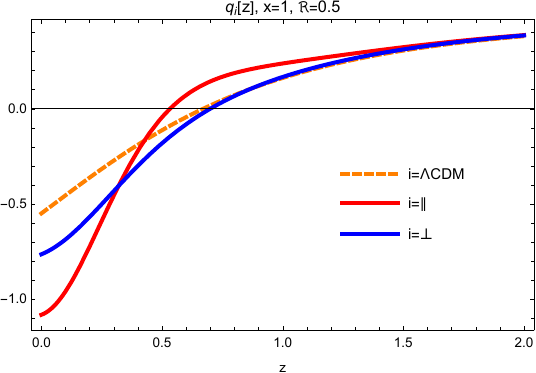}
\end{minipage}
\caption{On the left   the $z$ evolution of the decelerations  $q_{i}(z)$ (\ref{qzi})  for  $\Lambda$CDM model,   longitudinal and   transverse when $x=0.1,\,\Re=0.5$.
  On the right the same but with $x=1,\,\Re=0.5$.
}
\label{figQJz}
\end{figure}

\subsection{Drift}\label{drfrw}
\begin{figure}
\centering
\begin{minipage}{.5\textwidth}
  \centering
  \includegraphics[width=.9\linewidth]{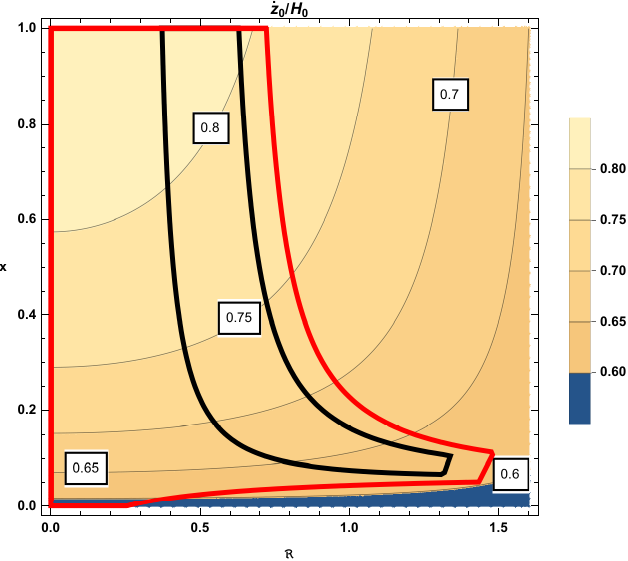}
\end{minipage}%
\begin{minipage}{.5\textwidth}
  \centering
  \includegraphics[width=.8\linewidth]{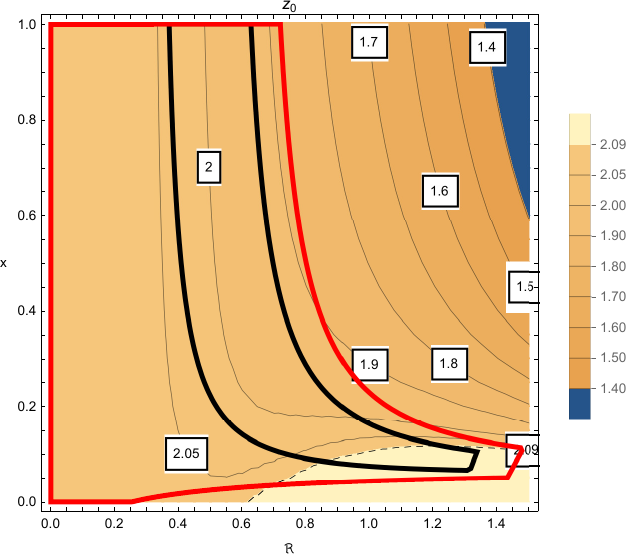}
\end{minipage}
\caption{On the left, we show the contourplot $\dot z_0/H_0$  (\ref{dz00}).
On the right, we present the contourplot of   the zero point $z_0$, where  $\dot z(z_{0})=0$. 
For   the $\Lambda$CDM model, we have   $\dot z_0/H_0=-q_0=0.55$ and
$z^{\Lambda CDM}_0=2.089$.
The black and red curves denote the $1-2\,\sigma$  regions from  fig.\ref{figOMk}.}
\label{Dzz0}
\end{figure}

\begin{figure}
\centering
\begin{minipage}{.5\textwidth}
  \centering
  \includegraphics[width=.8\linewidth]{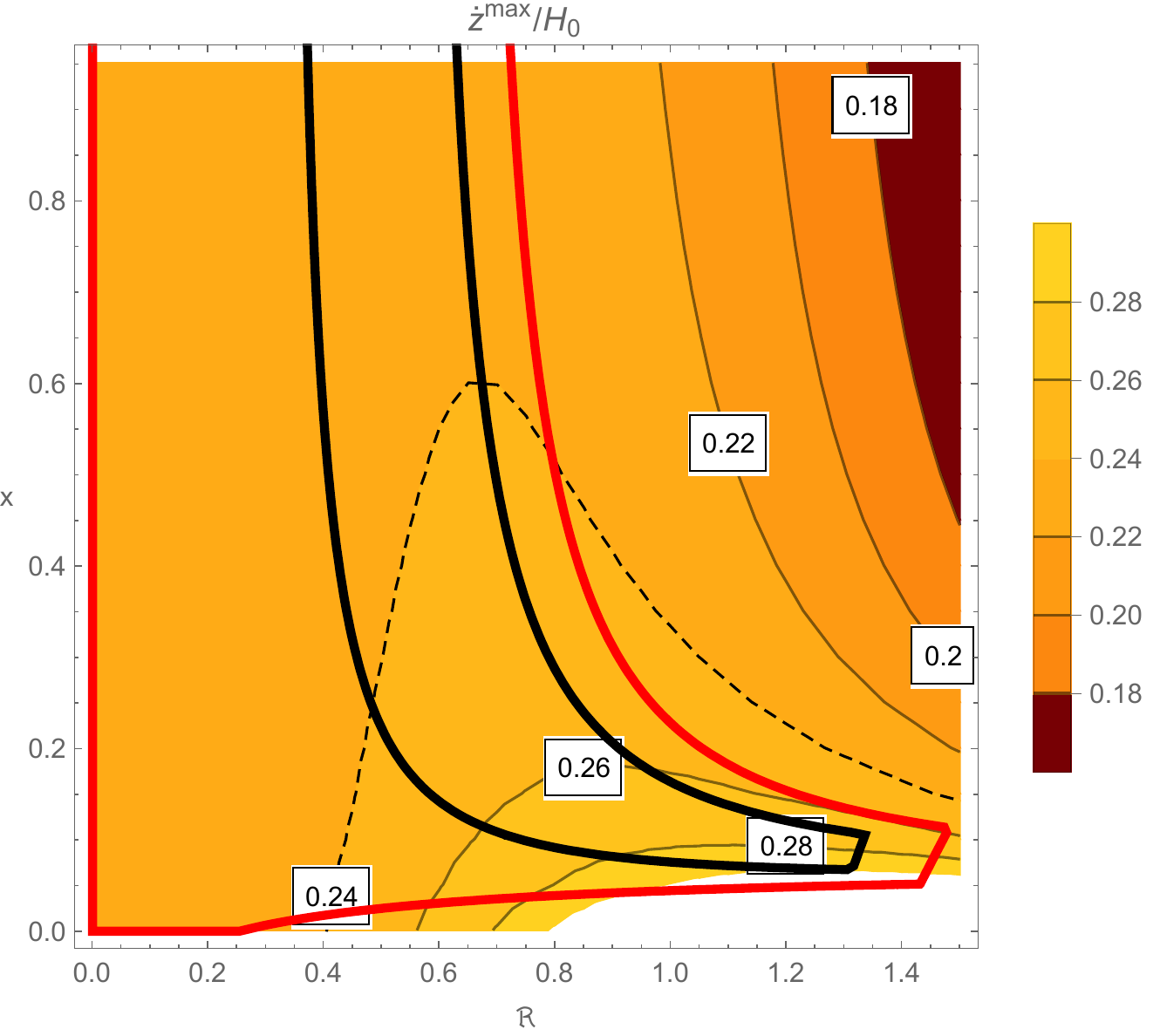}
\end{minipage}%
\begin{minipage}{.5\textwidth}
  \centering
  \includegraphics[width=.8\linewidth]{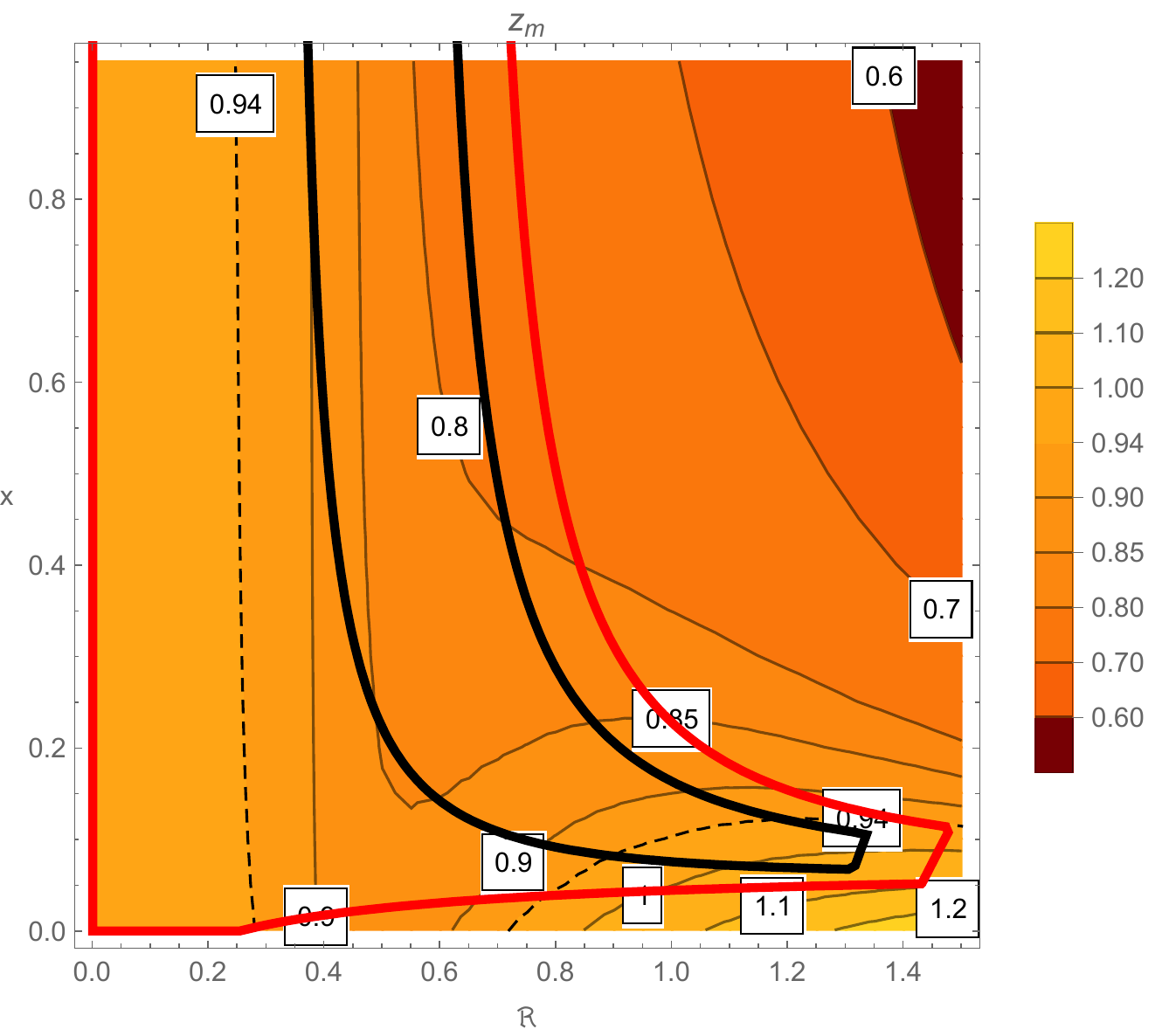}
\end{minipage}
\caption{On the left, we show the  contourplot of the maximum   $\dot z^{max}$ , which occurs  at $ z_{m}$. 
On the right,  we present the contourplot   of $z_m$, where  $\dot z(z_m)=\dot z^{max}$.
For   the $\Lambda$CDM model, we find $z_m=0.94$ and $\dot z^{max}/H_0=0.24$. 
The black and red curves denote the $1-2\,\sigma$  regions from  fig.\ref{figOMk}. }
\label{figDzmax}
\end{figure}

\begin{figure}
\centering
\begin{minipage}{.5\textwidth}
  \centering
  \includegraphics[width=.8\linewidth]{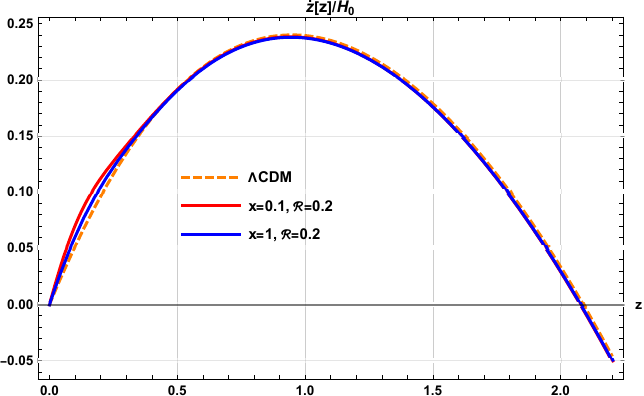}
\end{minipage}%
\begin{minipage}{.5\textwidth}
  \centering
  \includegraphics[width=.8\linewidth]{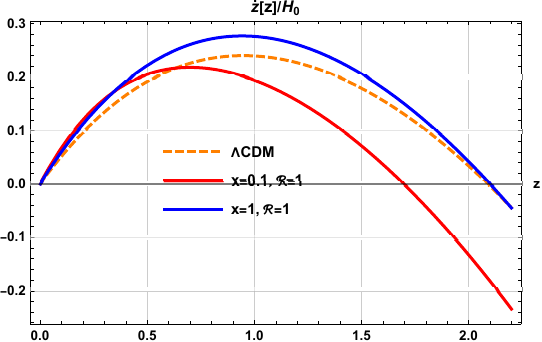}
\end{minipage}
\caption{The redshift evolution of $\dot z(z)/H_0$ is shown for the $\Lambda$CDM model (\ref{tzfrw}) and 
 our model (\ref{tzL}). 
 On the left, we plot the results  for $x=0.1$ and $x=1 $ with a small value of $\Re=0.2$. On the right,
  we show the same values of $x$  for a larger value of  $\Re=1$.}
\label{figDz0}
\end{figure}

In order to characterise the $z$ dependence of the drift function $\dot z(z)$ (\ref{tzL})  in the 
$x-\Re$ free parameter space we show various plots at different time scales.
\\
First of all, we start from the nearby value  $\dot z_0$ related to the expansion 
\be
\dot z(z)=  \dot z_0\,z+{\cal O}(z^2)\label{dz00}
\ee
 From fig.\ref{Dzz0} we see that, in the entire parameter space, we have systematically 
$\dot z_0>\dot z_0^{\Lambda CDM}=-q_\Lambda=0.55$ with a maximum of $\sim 0.83$ for large $x$ and small $\Re$. 
\\
To further characterise the  $\dot z$ evolution we note  two crucial features (see fig.\ref{figDz0}):
\begin{itemize}
\item A maximum around $z_{m}\sim1$ that we denote as
$\dot z^{max}$  (i.e. $\dot z^{max}=\dot z(z_m)$).
\item
 A zero point where  $\dot z(z_{0})=0$, that happens near  $ z_{0}\sim 2$.
\end{itemize}
In particular for the $\Lambda$CDM model, we have $z^{\Lambda CDM}_{m}=0.94$ and
$\dot z^{max}_{\Lambda CDM}/H_0=0.24$ and the zero point at $z^{\Lambda CDM}_0=2.09$.
In fig.\ref{figDzmax}   we contourplot the values of $\dot z^{max} /H_0$ (on the left)   and the value of $z_m$ (on the right) as a function of  $x-\Re$.
We observe that $\dot z^{max} /H_0$ is slightly larger than in the $\Lambda$CDM model in the preferred   region of parameter space, while in general, $z_m$ is smaller that the  $\Lambda$CDM value. This implies that the maximum drift occurs earlier in our model.
Finally, in fig.\ref{figDz0}, we show the $z$ evolution of the drift (\ref{tzL}) for two representative
 values in the  $x-\Re$ parameter space.
In summary, we find in our model, the redshift drift starts with a steeper slope
 near the origin, reaches its maximum earlier than in the  $\Lambda$CDM model,
 and attains a maximum value that is only slightly larger (at most  15 \%)  than the  $\Lambda$CDM model. 
 Afterward, the drift returns to zero earlier than in the $\Lambda$CDM model.
    
 
     \section{Conclusions}
  We analysed the dynamics of a cosmological model with  
    a Locally Rotationally  Symmetric  spacetime coupled with a specific perfect fluid.
   The key feature of this   fluid is the presence of a   space-dependent CC and, simultaneously,   a DM component, both exhibiting    the same exponential suppressing profile for the inhomogeneities  (\ref{le1}).
  A theoretical study of such a fluid is presented in \cite{Comelli:2023otq}, where only the space-dependent CC component was considered.
  Here, we skip  the detailed theoretical aspects  and instead   postulate a perfect fluid EMT (\ref{emt}), where the energy density and pressure contains  both DM and DE component (\ref{rho}). 
  Using a Lema$\hat{\rm i}$tre metric in comoving coordinates, we show that the Einstein eqs 
  (\ref{eqF}, \ref{eqY},  \ref{eqX}) are solvable, modulo some space-dependent functions that arise from  boundary conditions.
  A peculiarity of such a models, where the homogeneity of the spacetime is broken, is the presence of
  multiple independent expansion rates: the longitudinal   ${\cal H}_{\parallel}$
   (\ref{hl}), the  transverse  ${\cal H}_{\perp}$ (\ref{eqY}), and  the volume expansion rate ${\cal H}_{\theta}$ (\ref{Htheta}).
    In a FRW universe, these rates collapse into a single function,
   whereas  in a LTB spacetime, only two of them remain independent (\ref{Hs}).
   \\
    In Chapter \ref{geolc}, we derive the null geodesic  eqs    and the variation of the redshift $z$ along the light path (\ref{geol}).
    In Chapter \ref{dri}, we present the explicit eqs for the  redshift drift,  extended to the Lema$\hat{i}$tre metric (\ref{tzL}).
   In Chapters \ref{near} and  \ref{pertf}, we  solve  
     both the Einstein evolution eqs and the light geodesic paths in two different approximation schemes.
  \\
  In Chapter \ref{near}, we use a space Taylor expansion around the center and compute, order by order,   various cosmographic observables,  such as the decelaration parameter $Q_0$ (\ref{q0}) and the jerk $J_0$ (\ref{j0}). In eq.(\ref{dz0}), we provide the leading ${\cal O}(z)$ correction to the redshift drift (the second order ${\cal O}(z^2)$ corrections are presented in Appendix \ref{driftz}).
We establish a  consistency relationship  between the leading ${\cal O}(z)$ corrections to the Hubble rates and the redshift drift   (\ref{Hr}).
  \\
  In Chapter \ref{pertf}, we expand around FRW solution and treat all terms that break   homogeneity as perturbations (see (\ref{lmne})).
    The perturbed metric is derived from eqs (\ref{fxr}, \ref{xfr}, \ref{rfx}). 
       The solutions to the null geodesic eqs are  provided in Chapter \ref{lux} (\ref{arfrw}, \ref{eqa1}, \ref{eqr1}).  This allows us to  match the model  with the FRW external space,  where the initial densities $\Omega_i$ get {\it renormalised} at large distances to $\Omega_i^{eff}$ (\ref{Op}).
   We impose the Planck constraints \cite{Planck} on     $\Omega_i^{eff}$ to obtain bounds on the parameters space,  specifically for $x$, the fraction of DM versus DE (\ref{parl}), and the size of the bubble $\Re$ (\ref{le1}, \ref{leh1}). These bounds are illustrated in fig.\ref{figOMk}. 
   Notably, we find, at 1-$\sigma$ level, an absolute bound of   $x\gtrsim 0.07$ and also  
   $  \Re \lesssim 1.3$, further for $x\gtrsim 0.3$ the range of  $\Re$ is getting narrowed: $0.4\lesssim \Re\lesssim 0.6$. 
   So a pure DM profile with $x=0$ will be excluded at this level.
   At 2-$\sigma$ level, the above bounds get relaxed  however
   an upper bound of $\Re \lesssim 0.78$ for a DE dominated bubble (large $x\gtrsim 0.3$ values) remains. A small bubble of DM with $x=0$ is allowed for $\Re\lesssim 0.25 $.
    In fig.\ref{figH}, we show  the evolution of the various Hubble parameters with redshift for both large and small $x$ values at fixed $\Re=0.5$. 
  \\
   To extract more physical insights, we study the  evolution of the effective  equation of state 
   as a function of  $z$ for  different Hubble functions (\ref{wzi}). 
   At the origin  ($z\to0$), the effective eq.of state along the light of sight $w_\parallel(0)$ is always smaller that the transverse one  $w_\perp(0)$ with   minimum values of $w_\parallel(0)^{Min}\sim-1.07$ and $w_\perp(0)^{Min}\sim-0.69$. 
   During the $z$ evolution (see fig.\ref{figWz}), we observe a crossing where 
   $w_\parallel(z)$  becomes larger than $w_\parallel(z)$ followed by a gradual approach to the $\Lambda$CDM equation of state, $w_{\Lambda CDM}(z)$.
  \\
  For   the cosmographic deceleration parameter $Q_0$ and jerk $J_0$, we find   a minimum of
   $Q^{min}_0=-1.24$, to be compared with $-0.55$ in the $\Lambda$CDM, and a maximum of
  $J_0^{max}=3.75$ (for very large $\Re$ values and $x\to 1$),  compared with 1, with no lower bound for negative $J_0$ values. 
  For   the $z$ evolution of the $q_i(z)$ functions, we find $q_\parallel(z)<q_\perp(z)$ for small $z$ ($z < \Re$), and  around $z\sim \Re$,   $q_\parallel(z)>q_\perp(z)$, eventually merging asymptotically  with $q_{\Lambda CDM}(z)$.
  \\
  For   the drift, we compute  the nearby values, i.e.,  the leading term $\dot z_0$ (where 
  $\dot z\sim \dot z_0\,z+{\cal O}(z^2)$) as a function of the   free parameters $x-\Re$, as
  shown in the left panel of   fig.\ref{Dzz0}.
  The typical shape of the $\dot z(z)$ curve  resembles  a parabola, crossing the origin
   (see fig.\ref{figDz0}). 
  The "vertex" corresponds to the maximum of this function, which occurs  at $z_m$,  where 
  $\dot z(z_{m})=\dot z^{max}$.
  Another interesting point is the $z_0$ coordinate of the second null  point, where $\dot z(z_0)=0$.
  All the values of  $\dot z^{max}$, $z_m$ and $z_0$ are functions of the $x-\Re$ parameters,
   which we plot   in  fig.\ref{Dzz0} and fig.\ref{figDzmax}.
  \\
  In summary, we studied some of the main features of this exotic model, which,
   if the  $H_0$ crisis is confirmed, could provide valuable insights.
   The   choice of a single exponential behaviour for  the perturbations (\ref{le1}) is motivated  by 
   the economy of the free parameters ($x$ for  weighing DM versus DE, and $\Delta$ or $\Re$ for the size of the bubble). Note that the full amplitude of the effect  (DE+DM) is fixed by the observed $\Delta H_0$ (see eq. (\ref{dl})).
   Ultimately, a functional analysis that accommodates all present observables would provide full flexibility due to the arbitrariness of space-dependent functions.

\begin{appendices}
\section{ {1+3} and 1+1+2 Formalism}
For review/applications of the {1+3} and the 1+1+2  formalism see \cite{Ellis:1998ct} and \cite{Clarkson}.
\label{AppA}
\begin{itemize}
\item{1+3} Formalism
\\
The spacetime is split  into time and space relative
to a fundamental observer, represented by the timelike unit vector field $u^\a$, which corresponds 
to the observer’s 4-velocity. 
In this framework, the  1+3 covariant threading irreducibly
decomposes any 4-vector into a scalar part parallel to $u^\a$ and a 3-vector part orthogonal to $u^\a$.
Furthermore, any second rank tensor is covariantly and irreducibly split into scalar, 3-vector
and projected symmetric trace-free 3-tensor components.
The projection tensor onto the metric of the 3-space (S) orthogonal to $u^\a$ is given by
\bea
h_{\a\b}=g_{\a\b}+u_{\a}\;u_{\b},\qquad h^\a_\a=3
\eea
The covariant derivative of a scalar function $f$ projected along $u^\a$ is so defined
\bea
\dot f=u^\a\nabla_\a f
\eea
We can decompose the covariant derivative of $u^\a$ orthogonal to $u^\a$ giving
\be
\nabla_\a\;u_\b=-u_\a \;{\cal A}_\b+\frac{1}{3}\;\theta\; h_{\a\b}+\sigma_{\a\b}+\omega_{\b\a}
\ee
where we defined the following structures
\begin{itemize}
\item 4-Acceleration
\be
{\cal A}_\a\equiv u^\b\nabla_\b\;u_\a\equiv \dot u_\a 
\ee
\item Expansion
\be
  \theta \equiv \nabla^\a\;u_\a
 \ee
 \item Shear
 \be
 \sigma_{\a\b}=h^\lambda_\b\;\nabla_{(\lambda}\;u_{\rho )}\;h_\a^\rho-\frac{1}{3}\;\theta\; h_{\a\b} 
 \ee
 \item Vorticity
 \be
 \omega_{\a\b}=h^\lambda_\b\;\nabla_{[\lambda}\;u_{\rho ]}\;h_\a^\rho,\qquad \omega^\a=\e^{\a\b\g}\;\omega_{\b\g}
\ee
\end{itemize}
%
 \item {1+1+2} Formalism
 \\
 The 1 + 1 + 2 approach is based on a double foliation of the spacetime.
 Projection tensor which represents the metric of the 2-spaces   orthogonal to $u^\a$ and a spacelike ($v^\a\;v_\a =   1$) vector  $v^\a$.
\bea
N_{\a\b}=h_{\a\b}-v_{\a}\;v_{\b},\qquad N^\a_\a=2
\eea
  The hat-derivative is the derivative along the vector field $v^\a$ in the surfaces orthogonal to $u^\a$, for a scalar function $f$ we have the definitions
\bea
\hat f=v^\a\nabla_\a f 
\eea
We are now able to decompose the covariant derivative of $v^\a$ orthogonal to $u^\a$ giving
\bea
\nabla_\a\;u_\b=-{\cal A}\;u_\a\;e_\b+\left(\frac{\theta}{3}+\Sigma\right)\;v_\a\;v_\b+
\left(\frac{\theta}{3}-\frac{\Sigma}{2}\right)\;N_{\a\b}+\Omega\;\e_{\a\b}
\\
\nabla_\a\;v_\b=-{\cal A}\;u_\a\;u_\b+\left(\frac{\theta}{3}+\Sigma\right)\;v_\a\;u_\b+
 \frac{\phi}{2} \;N_{\a\b}+\xi\;\e_{\a\b}
\eea
\be
D_\a\;v_\b=v_\b \;a_\b+\frac{1}{2}\;\phi\; N_{\a\b}+\xi\;\e_{\a\b}+\zeta_{\a\b}
\ee
where
\begin{itemize}
\item Sheet expansion
\be
\phi=\delta_\a\;v^\a=D_\a\;v^\a
\ee
\item Twisting of the sheet (the rotation of $v^\a$)
\be
\xi=\frac{1}{2}\e^{\a\b}\;\delta_\a v_\b
\ee
\item Acceleration  of $v^\a$
\be a_\a=\hat v_\a
\ee
\item Shear of $v^\a$
\be
\zeta_{\a\b}=\delta_{\{\a}\;v_{\b\}}
\ee

\end{itemize}
Note the following identities and definitions  
  \bea
&&   v\cdot u=0,\quad v^2=1,\quad  v\cdot a=-u\cdot \dot v,\quad v\cdot\dot v=0
\\
  && {\cal A}={\cal A}^\a  v_\a =-u^\a\;\dot v_\a ,\quad \Sigma=v^\a\sigma_{\a\b} v^\b,\quad  
  \Omega=v^\a \o_{\a\b} v^\b,\quad
     \eea
\end{itemize}

\section{Small  \texorpdfstring{$z$}{Lg}  expansion for the redshift drift }
\label{driftz}

Starting from eq.(\ref{tzL}) we factorise the two contributions (the first is local $\dot z_L(z)$, the second is non local $\dot z_{NL}(z)$ in the $z$ space)
\be
\dot z(z)= \underbrace{H_0(1+z)-{\cal H}_{\parallel}(z)}_{\equiv \,\dot z_L(z)}+\underbrace{(1+z)\int^z_0\frac{1}{(1+z')^2}\frac{\partial_r\log(F\,{\cal H}_{\parallel})}{X}\,dz'}_{\equiv\,\dot z_{NL}(z)}
\ee
While it is straightforward the  small $z$  expansion   of the local contribution,  
 for $\dot z_1(z)$ we can use the following formula 
\bea
(1+z)\,\int^z_0 dz' \,f(z')\simeq f(0)\,z+\left(f(0)+\frac{f'(0)}{2}\right)\,z^2+{\cal O}(z^3)
\eea
The leading order ${\cal O}(z)$ corrections are given by\footnote{ The order ${\cal O}(z^2)$ corrections are given by
\bea\nonumber
\dot{z}_{L} (z)&\to&\frac{H_0 }{8}\, z^2\, \left(24\,\text{$\Delta $f}-4\, \Omega _{\kappa }+12 \,\Omega _m
   \left(\Omega _{\kappa }-2\, \Omega _{\chi }-1\right)+9\, \Omega _m^2-10\, \Omega _{\chi }+
  \right. \\&&\left. 4\,
   \left(\left(\Omega _{\kappa }-2\, \Omega _{\chi }\right){}^2-3\, \Omega _{\chi }\,
   R_2'(1)+R_2'(1){}^2\right)-8 \,R_2'(1)-4\, R_3'(1)+8\, R_2''(1)\right),
   \\ \nonumber
   \dot{z}_{NL}(z)&\to&\frac{H_0}{8} \,z^2\, \left(3 \,\Omega _{\chi } \left(2 \,\Omega _{\kappa }+3 \,\Omega _m+4\,
   R_2'(1)-2\right)-12 \,\Omega _{\chi }^2-4\, \left(6 \,\text{$\Delta
   $f}+\left(R_2'(1)-2\right) \,R_2'(1)-R_3'(1)+R_2''(1)\right)\right)
\eea
where $R_2'(1)$ and $R_2''(1)$ can be obtained from (\ref{eq1r}).
}
\bea\nonumber
&& \dot{z}_{L}(z)= z \,H\,\left( 1-\frac{3 \,\Omega _m}{2}-\Omega _{\kappa }+
2\, \Omega _{\chi } -R_2'(1)\right)+{\cal O}(z^2),\;\;
 \\&& \nonumber
\dot{z}_{NL}(z)\to z\,H\,
   \left(-\frac{3\,  \Omega _{\chi }}{2}+R_2'(1)\right)+{\cal O}(z^2),
 \\&&  
   \dot{z}(z)
   = H z \left(1-\frac{3\, \Omega _m}{2}-\Omega _{\kappa }+\frac{\Omega _{\chi
   }}{2}\right)  +{\cal O}(z^2)\label{redsh0}
\eea
The corrections  from $\dot z_L(z)$ and $ \dot z_{NL}(z)$ are of the same order as noted in   \cite{Koksbang:2022upf}, \cite{Codur:2021wrt}.

\end{appendices}

\end{document}